\documentclass[twoside,letterpaper,aps,prb,floatfix,twocolumn,showpacs,superscriptaddress,10pt,nobibnotes]{revtex4-1}
\usepackage{graphicx,times}
\usepackage{amssymb,amsmath}
\usepackage{multirow}
\usepackage{color}
\usepackage{epstopdf}
\usepackage[colorlinks=true, urlcolor=blue, linkcolor=red, citecolor=blue, pdftex]{hyperref}
\hypersetup{pdfstartview={FitV}, pdfpagemode = UseNone, pdfpagelayout=SinglePage, bookmarksnumbered={true}}

\newcommand{\braket}[2]{\langle #1|#2\rangle}
\newcommand{\bra}[1]{\ensuremath{\langle #1 |}}
\newcommand{\ket}[1]{\ensuremath{| #1 \rangle}}
\usepackage{ifthen}
\newlength{\abstractdiagwidth}
\newlength{\specdiagwidth}
\newcommand{\specdiag}[2][]{\parbox[c]{#2\specdiagwidth}{\includegraphics*[width=#2\specdiagwidth]{#1}}}%
\newcommand{\hon}[1]%
{%
\ifthenelse{\equal{#1}{2_1}}{\specdiag[Fig25]{154}}{}%
\ifthenelse{\equal{#1}{2_2}}{\specdiag[Fig26]{154}}{}%
\ifthenelse{\equal{#1}{4_2}}{\specdiag[Fig27]{191.5}}{}%
\ifthenelse{\equal{#1}{4_5}}{\specdiag[Fig28]{191.5}}{}%
\ifthenelse{\equal{#1}{4_6}}{\specdiag[Fig29]{191.5}}{}%
\ifthenelse{\equal{#1}{4_7}}{\specdiag[Fig30]{191.5}}{}%
}%
\begin{document}
\title{Phase Diagram of a Frustrated Quantum Antiferromagnet on the Honeycomb Lattice:\\
Magnetic Order versus Valence-Bond Crystal Formation}

\author{A.~F.~Albuquerque}
\affiliation{Laboratoire de Physique Th{\'e}orique, Universit{\'e} de Toulouse, UPS (IRSAMC), F-31062 Toulouse, 
France}
\affiliation{CNRS, LPT (IRSAMC), F-31062 Toulouse, France}

\author{D.~Schwandt}
\affiliation{Laboratoire de Physique Th{\'e}orique, Universit{\'e} de Toulouse, UPS (IRSAMC), F-31062 Toulouse, 
France}
\affiliation{CNRS, LPT (IRSAMC), F-31062 Toulouse, France}

\author{B.~Het\'enyi}
\affiliation{Max Planck Institut f\"ur Physik komplexer Systeme, D-01187 Dresden, Germany}
\affiliation{Institut f\"ur Theoretische Physik, Technische Universit\"at Graz, Petersgasse 16, A-8010 Graz, Austria}

\author{S.~Capponi}
\affiliation{Laboratoire de Physique Th{\'e}orique, Universit{\'e} de Toulouse, UPS (IRSAMC), F-31062 Toulouse, 
France}
\affiliation{CNRS, LPT (IRSAMC), F-31062 Toulouse, France}

\author{M.~Mambrini}
\affiliation{Laboratoire de Physique Th{\'e}orique, Universit{\'e} de Toulouse, UPS (IRSAMC), F-31062 Toulouse, 
France}
\affiliation{CNRS, LPT (IRSAMC), F-31062 Toulouse, France}

\author{A.~M.~L\"auchli}
\email{aml@pks.mpg.de}
\affiliation{Max Planck Institut f\"ur Physik komplexer Systeme, D-01187 Dresden, Germany}

\date{\today}

\begin{abstract}
We present a comprehensive computational study of the phase diagram of the frustrated $S=1/2$ Heisenberg
antiferromagnet on the honeycomb lattice, with second-nearest $(J_2)$ and third-neighbor $(J_3)$ couplings.
Using a combination of exact diagonalizations of the original spin model, of the Hamiltonian projected into the nearest neighbor 
short range valence bond basis, and of an effective quantum dimer model, as well as a self-consistent cluster mean-field theory,
we determine the boundaries of several magnetically ordered phases
in the region $J_2,J_3\in [0,1]$, and find a sizable magnetically disordered region in between. We characterize part of 
this magnetically disordered phase as a {\em plaquette} valence bond crystal phase. At larger $J_2$, we locate a sizable region in 
which {\em staggered} valence bond crystal correlations are found to be important, either due to genuine valence bond crystal ordering or 
as a consequence of magnetically ordered phases which break lattice rotational symmetry. 
Furthermore we find that a particular parameter-free Gutzwiller projected tight-binding wave function
has remarkably accurate energies compared to finite-size extrapolated ED energies along the transition line from conventional N\'eel to {\em plaquette} VBC phases, 
a fact that points to possibly interesting critical behavior - such as a deconfined critical point - across this transition.  We also comment on the relevance of this spin model to model
the spin liquid region found in the half-filled Hubbard model on the honeycomb lattice. 
\end{abstract}

\maketitle

Magnetic frustration is a very appealing route to weaken or destroy
magnetic order, which can result in new phases of matter: these phases
can usually be classified and named according to the broken symmetry
(spin, lattice) if any, or they can belong to the spin-liquid zoo when
no symmetry is broken~\cite{balents:10}. The quest for a genuine
gapped spin-liquid in a spin-1/2 model with SU(2) symmetry and an odd number of
sites in the unit cell has started a long-time ago with the proposal
by Anderson~\cite{anderson:73} that the ground-state of the Heisenberg
model on the triangular lattice could be viewed as a superposition
of short-range valence bonds (VB), called a resonating
valence bond (RVB) state. For the specific example of the triangular lattice
it turned out later however that a magnetically ordered state is 
realized~\cite{Bernu1992}.  Up to now, there is still no firmly established spin-liquid 
ground-state with the aforementioned properties in a reasonably realistic SU(2) 
spin model, although there are potential candidates, such as the triangular lattice 
with ring exchange interactions~\cite{Misguich1998} or the Heisenberg model on the kagom\'e lattice~\cite{Yan2010}.
On the other hand, if one considers lattices with an even number of sites per unit cell, 
then Hasting's generalization~\cite{Hastings2004} of the Lieb-Schultz-Mattis theorem~\cite{Lieb1961} 
does not apply, and it is possible in principle to stabilize a magnetically disordered ground-state that
does not break any symmetry and has only trivial topological properties. One can think for instance 
of a Heisenberg model on a square bilayer lattice with strong interlayer exchange. The honeycomb lattice
is peculiar in this respect because no simple lattice-symmetry preserving deformation is known which would
lead to a gapped magnetic state~\footnote{This is possibly related to the fact that no symmetry-preserving 
tight binding model on the honeycomb lattice is known, which would gap out the Dirac cones.}.

\begin{figure}
  \begin{center}
    \includegraphics* [width=0.8\linewidth]{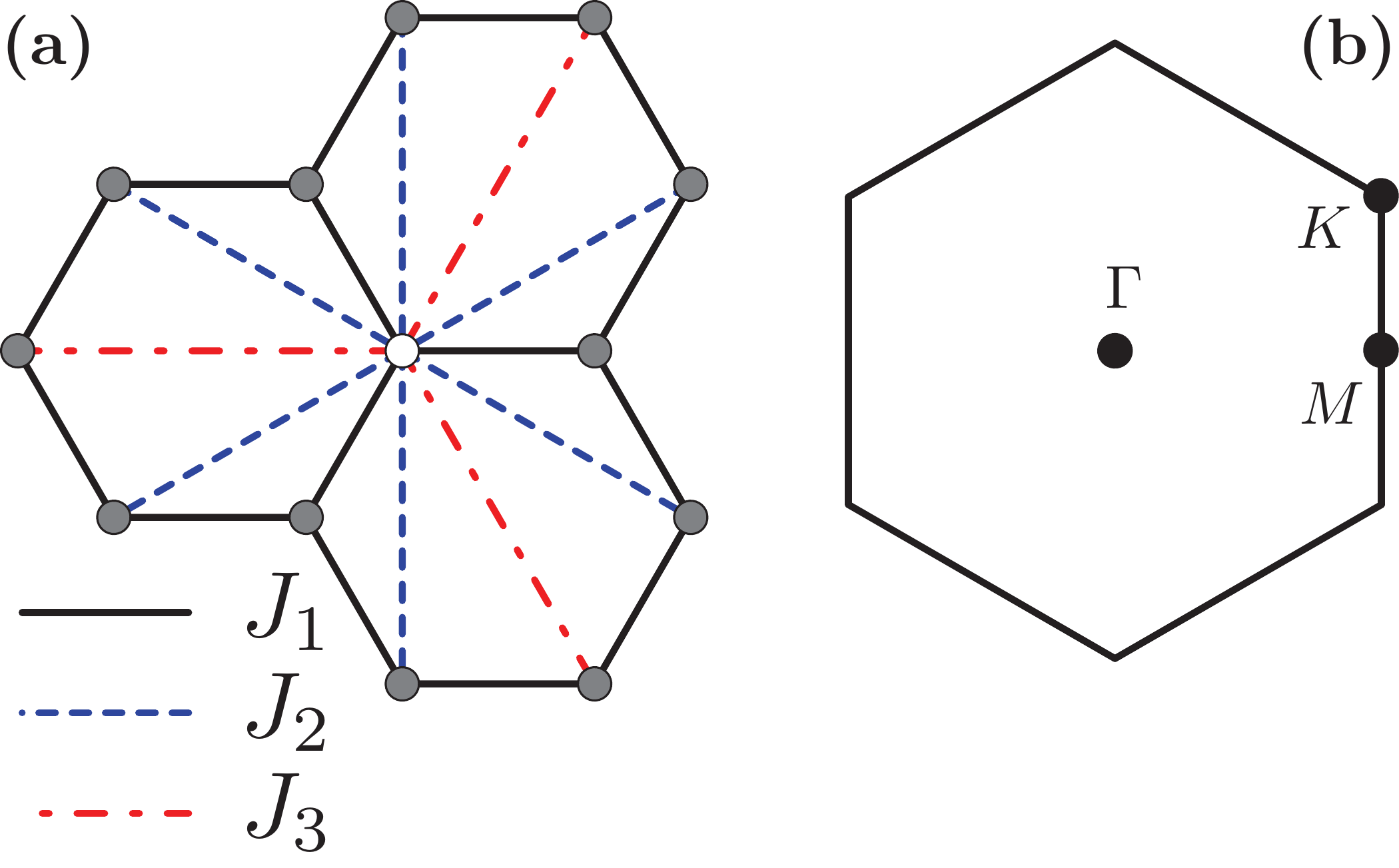}
   \end{center}
   \caption{(Color online) (a) Honeycomb lattice with the different spin exchange interactions considered in this paper; (b) corresponding Brillouin zone with relevant ${\bf k}$ points.}
  \label{fig:lattice}
\end{figure}
\begin{figure*}[t]
  \begin{center}
    \includegraphics*[width=0.9\linewidth]{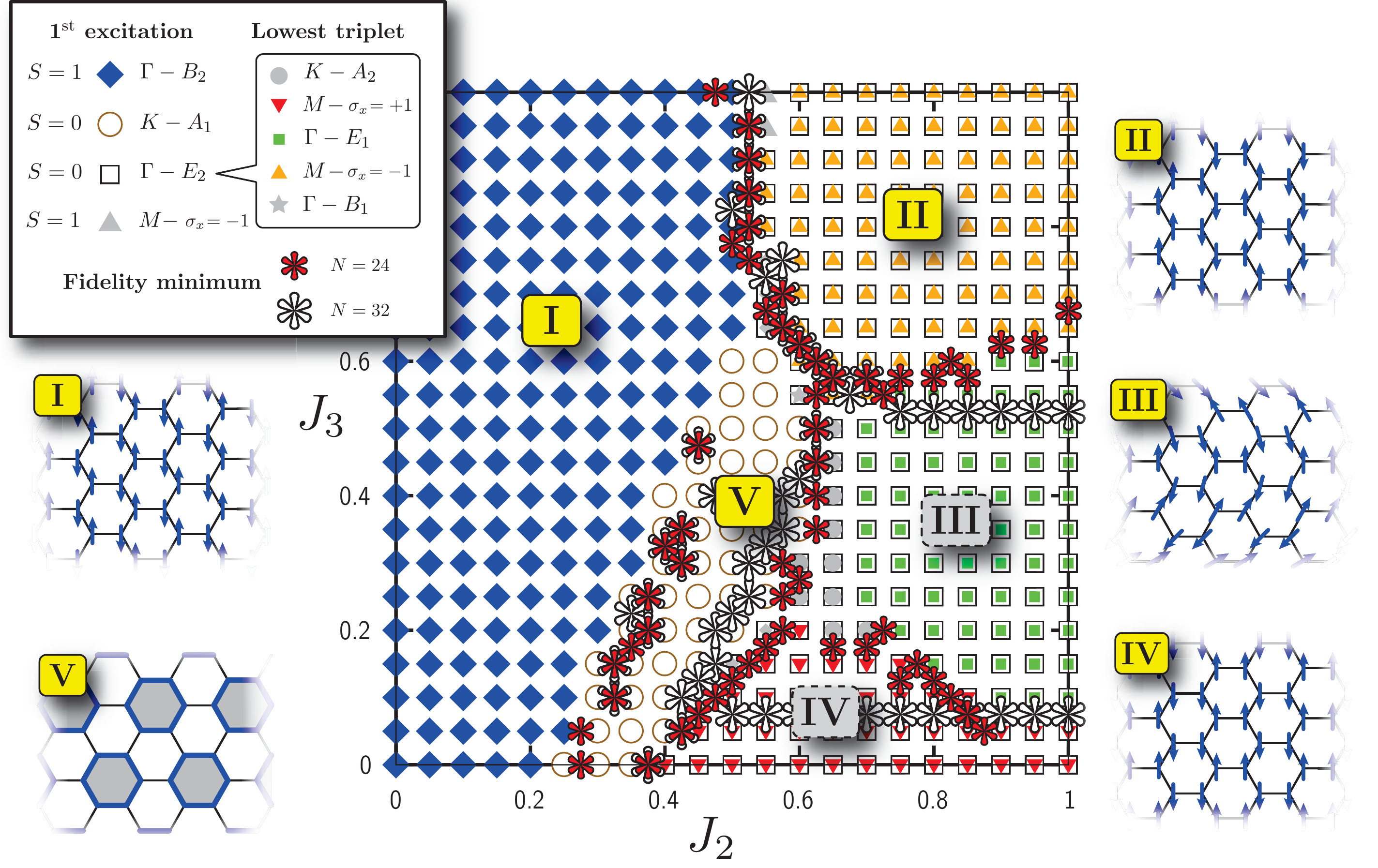}
   \end{center}
   \caption{(Color online) Phase diagram of the frustrated $S=1/2$ Heisenberg model honeycomb lattice
   in the region $J_2,J_3 \in [0,1]$, based on a combination of exact diagonalization results discussed in
   the main text. The 5 regions identified here correspond to: {\bf (I)} a N\'eel ordered phase with staggered 
   magnetization, {\bf (II)} a collinear magnetically ordered phase,
   {\bf (III)} One or several phases corresponding to short or long range ordered non-collinear magnetic order,
   {\bf (IV)} A different collinear magnetically ordered (or disordered) phase corresponding to phase (IV) in Ref.~\onlinecite{Fouet2001}
   and {\bf (V)} a magnetically disordered phase forming a {\em plaquette} valence bond crystal.  The five phases are sketched in
   the panels around the phase diagram.  Note that the phases highlighted in grey (III), (IV) show substantial finite size effects 
   and are therefore difficult to characterize precisely.
   }
  \label{fig:phase_diagram}
\end{figure*}

In recent years a promising new direction in the search for spin liquids has opened up, focusing on
the behavior of insulating phases upon approaching the Mott insulator-metal transition. In
the half-filled triangular lattice Hubbard model a picture with a spin bose metal spin liquid phase
sandwiched between the metallic phase at small $U/t$ and the magnetically ordered N\'eel phase
at large $U/t$ has emerged~\cite{Morita2002,Yoshioka2009,Motrunich2005,Sheng2009,Yang2010}. 
It has been recognized that this spin liquid phase can be understood in terms of a pure spin model, 
where the rising charge fluctuations are cast into an increasingly complex spin Hamiltonian beyond the 
Heisenberg model~\cite{Motrunich2005,Yang2010}. A second striking example of a spin liquid located
between a magnetically ordered phase and a (semi)metal has recently been uncovered in the half-filled 
Hubbard model on the honeycomb lattice~\cite{Meng2010}. Such spin liquid phase is reported
to have a small spin gap and no appreciable correlations of any kind. 

This exciting finding leads us to the natural question of whether this spin liquid phase on the honeycomb
lattice can also be described within a pure $S=1/2$ spin model, despite the vicinity of the insulator to 
semimetal transition. A high order derivation of the corresponding
spin model is in progress~\cite{Yang2011}, but the typical value of the expansion parameter $t/U \sim 0.25 $
relevant for the spin liquid phase renders this task more challenging in comparison to the triangular lattice, 
where a typical value for the spin liquid regime is about $t/U \sim 0.11$. In the absence of an accurate
prediction for a relevant spin model, we start by exploring the effect of the next-to-leading order 
correction to the nearest neighbor Heisenberg model, which is a second neighbor Heisenberg coupling
$J_2$ arising at fourth order in $t/U$. We thus consider in the following a frustrated $S=1/2$ Heisenberg Hamiltonian
on the honeycomb lattice, where we also include a third neighbor coupling $J_3$ for completeness.

The honeycomb (hexagonal) lattice Hamiltonian [see Fig.~\ref{fig:lattice}(a)] reads:
\begin{equation}
{\cal H}=J_1 \sum_{\langle i,j\rangle} \mathbf{S}_i\cdot \mathbf{S}_j 
+J_2 \sum_{\langle\langle i,j\rangle\rangle} \mathbf{S}_i\cdot \mathbf{S}_j
+J_3 \sum_{\langle\langle\langle i,j\rangle\rangle\rangle} \mathbf{S}_i\cdot \mathbf{S}_j .
\label{eqn:HeisenbergHamiltonian}
\end{equation}
In this paper we focus solely on antiferromagnetic interactions $J_a\ge0$, set $J_1=1$ 
and restrict ourselves to the window $J_2,J_3 \in[0,1]$. Aspects of this frustrated model have been explored
previously in the literature, based on spin wave theory~\cite{Rastelli19791,Fouet2001,Mulder2010,Ganesh2010},
a nonlinear sigma model treatment~\cite{Einarsson1991}, Schwinger boson 
approaches~\cite{Mattsson1994,Cabra2010} and exact diagonalizations~\cite{Fouet2001,Mosadeq2010}. Note also that a similar frustrated model gives rise to a rich phase diagram on the square lattice~\cite{Mambrini2006,Reuther2011}.

In this work we thoroughly explore the phase diagram in the considered window, based on a combination
of exact diagonalizations (EDs) of the spin model (up to 42 spins), EDs in the nearest-neighbor valence bond
(NNVB) subspace (up to 96 spins), EDs of an effective quantum dimer model (QDM) (corresponding to
up to 126 spins), complemented by a self-consistent cluster mean field theory (SCMFT) and the study of a fully 
projected Gutzwiller wave function of the half-filled honeycomb tight binding "Dirac sea".

The key finding of our work is the presence of a sizable magnetically disordered region adjacent to the
well studied N\'eel phase of the unfrustrated honeycomb Heisenberg model. We identify a large part of this region as
a {\em plaquette} valence bond crystal (VBC). Interestingly we find evidence (within the ED realm) for a possibly continuous
phase transition between the N\'eel phase and a {\em plaquette} valence bond crystal. In addition the energy and
some of the key correlations of the frustrated spin model in the transition region are well captured by a simple
Gutzwiller projected (GP) "Dirac sea" wave function. These findings raise the possibility of a continuous 
quantum phase transition beyond the Ginzburg Landau paradigm in this honeycomb lattice spin model. 

The outline of the paper is as follows: we start by giving a quick overview of the phase diagram in section~\ref{sec:PhDiag}.
Then the magnetically ordered phases are located using a SCMFT in section~\ref{sec:SCMFT}.
Next we study the spin model using EDs in section~\ref{sec:ED}, followed by EDs in the
NNVB subspace and EDs of a QDM which are presented in section~\ref{sec:EDVB}. We close with a discussion
and conclusion section~\ref{sec:conclusion}. In the appendices we discuss the properties of the Gutzwiller projected "Dirac sea"
(App.~\ref{sec:prop_gutzwiller}), present the derivation of an effective quantum dimer model from the Hamiltonian projected into the NNVB subspace 
(App.~\ref{sec:derivationeffQDM}), compare the energies and the finite size behavior of the NNVB versus the QDM approach 
(App.~\ref{sec:comp_NNVB_QDM}) and derive the expected correlation functions in model valence bond crystal states 
(App.~\ref{sec:PureStates}).

\section{Overview of the Phase Diagram}
\label{sec:PhDiag}

We start by summarizing the main result of this paper, the phase diagram of the frustrated $S=1/2$ Heisenberg Hamiltonian 
Eq.~\eqref{eqn:HeisenbergHamiltonian}
in the considered parameter window  displayed in Fig.~\ref{fig:phase_diagram}. The phase diagram emerges from a 
combination of different information extracted from ED of the spin model: 

(i) For a first analysis without further input we have investigated the structure of the fidelity $f$, i.e. the overlap between ground-states (GS) obtained for different parameters
\footnote{see e.g. Ref.~\protect{\onlinecite{Gu2010}} for a recent review on fidelity approaches to quantum phase transitions}: 
\begin{equation} 
f(J_2,J_3|J_2',J_3')=|\langle \mathrm{GS}(J_2,J_3 ) | \mathrm{GS}(J_2',J_3')\rangle|,
\end{equation}
for consecutive points along the two directions of the $(J_2,J_3)$ plane using a grid spacing of $0.05$.
Local minima of $f$ in the directions of $J_2$ and $J_3$ of $f$ are indicated by star symbols for both the $N=24$ and
$N=32$ samples. These dips already give a first impression of some phase boundaries in the phase diagram.

(ii) In addition we highlight the quantum numbers of the lowest excited state and - if not a triplet already - the quantum numbers  
of the lowest triplet, both for $N=24$.  The basic idea is that for sufficiently large systems the quantum numbers of the low 
energy spectrum are characteristic of the respective phases, and can thus be used to chart a phase diagram if used with care.
For a detailed discussion of the expected low lying energy levels in the different phases we refer to subsection~\ref{subsec:lowest_excitation}.

Based on these and further results to be discussed later, the following phases are identified: 
{\bf (I)} a N\'eel ordered phase with a finite staggered magnetization, located around the unfrustrated point
$J_2{=}J_3{=}0$; 
{\bf (II)} a magnetically ordered collinear phase corresponding to the classical phase (II) in Ref.~\onlinecite{Fouet2001} 
arising at combined large $J_2$ and $J_3$;
{\bf (III)} one or several phases corresponding to short-range or long-range non-collinear magnetic order, resulting from the $J_1,J_3$ 
coupling of two decoupled triangular lattices in the large $J_2$ limit;
{\bf (IV)} a collinear magnetically ordered phase (corresponding to phase (IV) in Ref.~\onlinecite{Fouet2001})
or a {\em staggered} dimer phase (also called lattice nematic in Ref.~\onlinecite{Mulder2010}).
And finally {\bf (V)} a magnetically disordered phase forming a {\em plaquette} valence bond crystal. 

Analyzing the magnitude of the fidelity dips it seems likely that the transition from phase (I) to (V) is continuous (corresponding to a faint feature
in the fidelity), while the transitions from (I) to (II) and (V) to (II), (III) and (IV)  seem to be of first order because of strong avoided level crossings 
observed on the clusters considered. The topology of the phase diagram and the nature of the phases in the regions (III) and (IV) display strong 
finite size effects and require further investigations beyond the scope of this work. We note that phases (II),(III) and (IV) exhibit long range
{\em staggered} dimer order also in the case of long-range magnetic order, because in the magnetically ordered phases one of the nearest neighbor (NN) bond energies 
is different from the other two.

We now proceed to a self consistent cluster mean field treatment which is well suited to detect various magnetically ordered phases. 

\section{Self consistent cluster mean-field theory}
\label{sec:SCMFT}

The very same frustration accounting for the rich physics exhibited by the here considered model [Eq.~\eqref{eqn:HeisenbergHamiltonian}]
also adds enormous complexity to the task of determining its properties.  In this context, approximate approaches can be valuable and
employed in obtaining ``draft phase diagrams" that may guide subsequent application of more accurate techniques. The so-called
self-consistent cluster mean-field theory (SCMFT)\cite{zhao:07, hassan:07} is a tool particularly well suited to this task. In a nutshell,
SCMFT consists in diagonalizing the Hamiltonian under investigation on small clusters that, besides including actual in-cluster couplings,
so that quantum fluctuations at the local level are partially taken into account, are also coupled to mean fields that are to be determined
self-consistently. This technique has been shown to considerably improve upon more conventional mean-field approaches for the case of hard-core
bosons on the triangular lattice\cite{hassan:07} and, more recently, to yield results that compare well with the ones from more sophisticated
techniques when applied to an effective model for a frustrated antiferromagnet.\cite{albuquerque:10b}

\begin{figure*}
  \begin{center}
    \includegraphics*[width=0.9\linewidth]{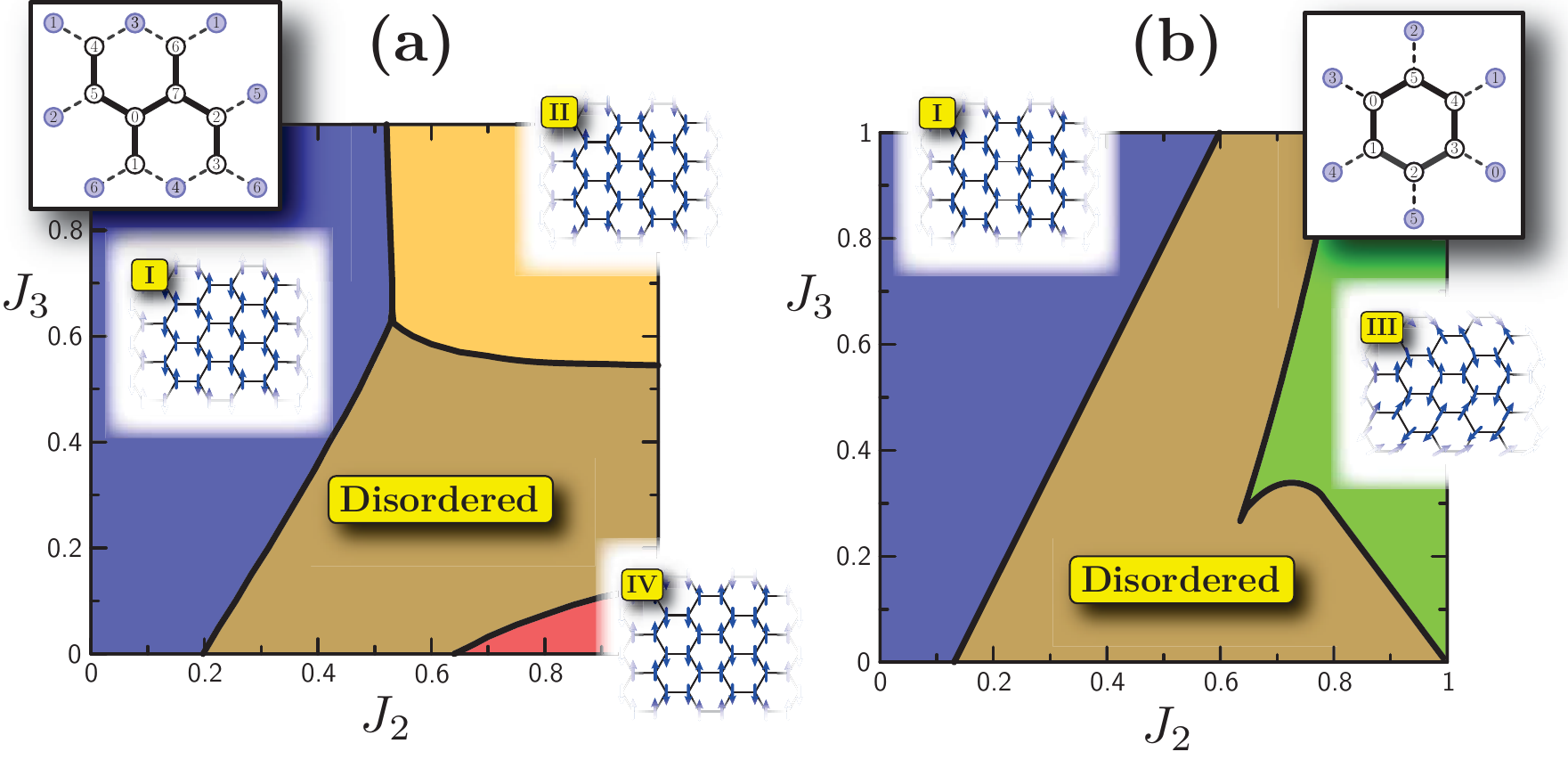}
   \end{center}
   \caption{(Color online) Phase diagram obtained from SCMFT as applied to the $N=8$ (a)  and $N=6$ (b) clusters.
   In the region labeled as ``disordered" no $SU(2)$-broken magnetic phases are obtained from the SCMFT procedure for
   the considered cluster. Insets: clusters employed in our SCMFT calculations. Thick
   lines connecting open circles represent in-cluster couplings and dashed lines
   coupling to mean fields (only $J_1$ interactions are depicted).}
  \label{fig:MF_phdiag}
\end{figure*}

In applying SCMFT, we consider the clusters comprising $N=6$ and $8$ sites depicted in the insets of Fig.~\ref{fig:MF_phdiag}.
We split the Hamiltonian Eq.~\eqref{eqn:HeisenbergHamiltonian} according to
\begin{equation}
    \hat{{\mathcal H}} = \hat{{\mathcal H}}_{\rm in} + \hat{{\mathcal H}}_{\rm MF}~.
  \label{eq:H-scmft}
\end{equation}
$\hat{{\mathcal H}}_{\rm in}$ accounts for in-cluster couplings,
\begin{equation}
    \hat{{\mathcal H}}_{\rm in} =  \sum_{\left\langle i,j \right\rangle} \mathbf{S}_{i} \cdot \mathbf{S}_{j} +
                               J_{2}  \sum_{\left\langle \left\langle i,j \right\rangle \right\rangle} \mathbf{S}_{i} \cdot \mathbf{S}_{j} +
                     J_{3}  \sum_{\left\langle \left\langle \left\langle i,j \right\rangle \right\rangle \right\rangle} \mathbf{S}_{i}
                     \cdot \mathbf{S}_{j} ~,
  \label{eq:in-cluster}
\end{equation}
and is treated in an exact way. ${\left\langle i,j \right\rangle}$, ${\left\langle \left\langle i,j \right\rangle \right\rangle}$ and
${\left\langle \left\langle \left\langle i,j \right\rangle \right\rangle \right\rangle}$ respectively denote nearest-, second-nearest-
and third-nearest-neighbor {\em in-cluster} sites (open circles in Fig.~\ref{fig:MF_phdiag}, where in-cluster NN bonds are represented
by thick lines). Couplings to the mean fields are included in ${\mathcal H}_{\rm MF}$ that reads
\begin{equation}
    \hat{{\mathcal H}}_{\rm MF} =  \sum_{{\left[ i,j \right]}} \mathbf{S}_{i} \cdot \langle \mathbf{S}_{j} \rangle +
        J_{2}  \sum_{{\left[ \left[ i,j \right] \right]}} \mathbf{S}_{i} \cdot \langle \mathbf{S}_{j} \rangle +
       J_{3}  \sum_{{\left[ \left[ \left[ i,j \right] \right] \right]}} 
       \mathbf{S}_{i} \cdot \langle \mathbf{S}_{j} \rangle ~.
  \label{eq:mean-field}
\end{equation}
Here, $\mathbf{S}_{i}$ denotes a spin-operator attached to an in-cluster site $i$ and is coupled to the mean-field given by the expectation value
$\langle \mathbf{S}_{j} \rangle$ at the ``across-the-boundary site" $j$ [light-filled circles in Fig.~\ref{fig:MF_phdiag}], for nearest
(${\left[ i,j \right]}$; dashed lines in Fig.~\ref{fig:MF_phdiag}), second-nearest (${\left[ \left[ i,j \right] \right]}$) and third-nearest
(${\left[ \left[ \left[ i,j \right] \right] \right]}$) neighbors. That is, one may see SCMFT as an exact diagonalization on a finite cluster with
periodic boundary conditions (PBC), where ``across the boundary" interactions are replaced by couplings to mean fields that are determined in a
self-consistent manner. One starts from a randomly chosen wave function and computes the mean fields $\langle \mathbf{S}_{j} \rangle$
at every site $j$, that are then used in setting ${\mathcal H}_{\rm MF}$. The cluster Hamiltonian Eq.~(\ref{eq:H-scmft}) is then diagonalized
and the so-obtained ground-state wave function is used in re-setting ${\mathcal H}_{\rm MF}$; computation proceeds until all $\langle
\mathbf{S}_{j} \rangle$ are converged and the existence of $SU(2)$-broken magnetic phases is signaled by non-vanishing mean fields, $\langle
\mathbf{S}_{j} \rangle \neq 0$.

In Fig.~\ref{fig:MF_phdiag} we present the resulting SCMFT phase diagrams obtained for
(a) an $N=8$ cluster and (b) an $N=6$ cluster~\cite{comment_scmft_01}.
For the $N=8$ cluster we first note the presence of two collinear magnetically ordered phases, labeled (I) and (II) in Fig.~\ref{fig:phase_diagram}.
These phases are also present in the classical version of the model, occupying roughly the same portion of the plane $(J_{2}, J_{3})$~\cite{Rastelli19791,Fouet2001}.
Furthermore, the phase labelled as IV in Fig.~\ref{fig:phase_diagram} --- also observed in
the classical case but only for $J_{3} < 0$,\cite{Rastelli19791,Fouet2001} --- occupies part of the region shown to support a spiral phase in Ref.~\onlinecite{Rastelli19791}.
This might be an interesting effect, where the {\em collinear} phase IV is stabilized for some $J_3 \ge 0$ (i.e. beyond the classical domain of stability) 
by quantum fluctuations. Note that a magnetically ordered phase of this type is also compatible with the pronounced {\em staggered} dimer pattern 
reported in previous ED studies~\cite{Fouet2001,Mosadeq2010} for $J_2 \gtrsim 0.4$, and the lattice nematic point of view~\cite{Mulder2010}.

In order to study finite size effects we apply SCMFT to an $N=6$ site cluster and present its phase diagram in
Fig.~\ref{fig:MF_phdiag}(b). First, we remark that this cluster [depicted in the inset of Fig.~\ref{fig:MF_phdiag}(b)] is not compatible with
both phases II and IV and that no solutions with $\langle \mathbf{S}_{j} \rangle \neq 0$ are encountered in some parts of the region stabilizing
these orderings for the $N=8$ site cluster [Fig.~\ref{fig:MF_phdiag}(a)]. Furthermore, the size of the region supporting N\'{e}el order is somewhat
reduced\cite{comment_scmft_02} in comparison with what is observed in Fig.~\ref{fig:MF_phdiag}(a): we suspect that this may be explained by the fact
that the Kekul\'{e}-like state with resonating valence bonds is particularly stable on the hexagon-shaped $N=6$ cluster, with the consequence that
the ``disordered" region is overestimated. More interestingly, and in contrast with what happens for the $N=8$ cluster, a spiral state (phase III)
is stabilized for large $J_{2}$. Such a state is adiabatically connected to the ground-state for $J_{2}/J_{1} \gg 1$, where Eq.~\eqref{eqn:HeisenbergHamiltonian}
decouples into two triangular lattices, each of which exhibits $120^{\circ}$ magnetic order.

Finally, we address the possible occurrence of non-magnetic phases for Eq.~\eqref{eqn:HeisenbergHamiltonian}. We notice the existence of an extended
region in $(J_{2}, J_{3})$ where vanishing mean-field solutions, $\langle \mathbf{S}_{j} \rangle = 0$, are obtained for both clusters considered 
in Fig.~\ref{fig:MF_phdiag}. Intersecting the magnetically disordered phases from both phase diagrams ($N=6$ and $N=8$)
we obtain a putative nonmagnetic region which roughly corresponds to the extent of phase V in the ED based phase diagram shown in Fig.~\ref{fig:phase_diagram}.

In the following we address the nature of this magnetically disordered region in more detail and determine some of the phase boundaries with higher accuracy
based on finite size extrapolated ED simulations.

\begin{figure*}[!ht]
\begin{center}
\includegraphics[width=0.7\linewidth]{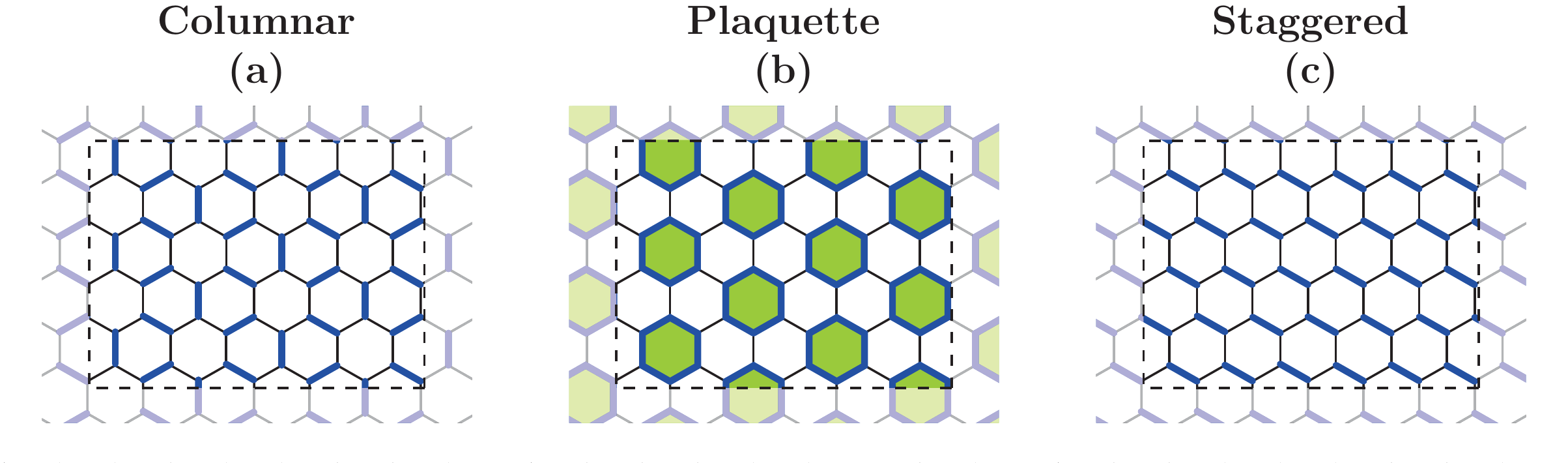}
\caption{(Color online) Pictorial representation of the three valence bond crystal candidate states discussed in this work. The {\em columnar} VBC
(a) is also called Read-Sachdev~\cite{Read1990} state in the literature, while the {\em staggered} dimer VBC is also known as "lattice nematic"~\cite{Mulder2010}.}
\label{fig:StateDefinitions}
\end{center}
\end{figure*}

\section{Exact diagonalization}
\label{sec:ED}

We now explore the phase diagram 
based on large scale ED in the $S^z$ basis of finite honeycomb lattice samples with $N=24,26,28,30,32$~\cite{Fouet2001,Mosadeq2010}$,34,36,38$\cite{RichterLNP2004} and $42$ sites.
The clusters with $N=24,30,36,42$ sites feature the two $K$ points in their Brillouin zone [c.f.~Fig.~\ref{fig:lattice}(b)], while $N=24,28,32,36$ contain one 
or several $M$ points. The clusters $N=24,26,32,38,42$ exhibit six-fold rotational symmetry.

We first study the low-lying energy spectrum in the full parameter region in subsection~\ref{subsec:lowest_excitation} in order to provide more information on the phase diagram shown
in Fig.~\ref{fig:phase_diagram}. Then we address the stability of the N\'eel phase (I) by calculating magnetic structure factors and energy scalings in subsection~\ref{subsec:neel_stability},
and close this section with a discussion of the nature of the dimer-dimer correlations in subsection~\ref{subsec:ed_dimercorrs}, supporting the presence of an
extended {\em plaquette} valence bond crystal phase. 

\subsection{Nature of the lowest excitation}
\label{subsec:lowest_excitation}

In compiling the phase diagram shown in Fig.~\ref{fig:phase_diagram}, one set of information was gathered from the quantum numbers of the low-lying excitations. The idea is that symmetry broken
phases must exhibit a specific set of low lying energy levels, which will allow the spontaneous symmetry breaking in the thermodynamic limit. In the case of $SU(2)$ symmetry breaking states, the 
appropriate structure is called ``tower of states" (TOS) and has been successfully used to identify magnetically ordered phases~\cite{Bernu1992}, as well as spin nematic phases~\cite{Lauchli2005}. The
finite size behavior of energy gaps is as follows: the levels belonging to the symmetry breaking tower states scale as $1/N$ with system size, while the spin-wave modes scale as $1/L\ $~\cite{Bernu1992,Lhuillier2001}.
For sufficiently large system sizes one should therefore detect only states belonging to the TOS manifold in the lowest part of the energy spectrum. In the case of discrete symmetry breaking - such
as for a VBC - a finite number of levels is expected to collapse rapidly (exponentially beyond a certain correlation length) onto the ground state. In each case, the quantum numbers of the
collapsing levels are determined by the nature of the order parameter, i.e. the broken symmetries, and generally they will be different for distinct phases (but not always). 
In the following we summarize the expected quantum numbers of low energy levels of several candidate phases (some of the results were presented earlier in Ref.~\onlinecite{Fouet2001}). Note that the quoted quantum numbers are given as appropriate for the $N=24$ sample, and the $C_{6v}$ point group is located at the center of a hexagon.

\begin{enumerate}
\item The N\'eel ordered phase (I) has a simple TOS structure with one level per total spin: all even spin sectors
belong to the $\Gamma$ A1 representation, while the odd ones appear in $\Gamma$ B2.
\item The magnetically ordered phase (II) has three levels per spin sector: the levels in the even spin sectors 
are found in $\Gamma$ A1 and the two-dimensional representation $\Gamma$ E2. The levels in the odd spin sectors
belong to the threefold degenerate $M$ momentum, with even (odd) parity for reflections along (perpendicular) to 
the $\Gamma -M$ axis.
\item The magnetically ordered phase (IV) has three levels per spin sector: the levels in the even spin sectors 
are found in $\Gamma$ A1 and the two-dimensional representation $\Gamma$ E2. The levels in the odd spin sectors
belong to the threefold degenerate $M$ momentum, with even parity for both reflections along and perpendicular to 
the $\Gamma -M$ axis.
\item A {\em columnar} (Read-Sachdev~\cite{Read1990}) [cf. Fig.~\ref{fig:StateDefinitions}~(a)] or {\em plaquette} VBC [cf. Fig.~\ref{fig:StateDefinitions}~(b)]  
has three collapsing singlet levels: one at $\Gamma$ A1 and a two-dimensional $K$ A1 representation. Note that these two VBCs
can not be distinguished based on energy level quantum numbers alone.
\item A {\em staggered} VBC [cf. Fig.~\ref{fig:StateDefinitions}~(c)] has three collapsing levels: $\Gamma$ A1 and $\Gamma$ E2 (2-dim representation).
\end{enumerate}

It is interesting to note that the level crossing of the excited states quantum numbers shown in Fig.~\ref{fig:phase_diagram} match quite well the dips in the fidelity (which is a ground state
observable). The quantum numbers carry however more information and allow to label the quantum phases roughly, before studying them in more
detail using correlation functions, as we will do in the following.

\subsection{Stability of the N\'eel phase}
\label{subsec:neel_stability}
\begin{figure}
\begin{center}
\includegraphics[width=0.98\linewidth]{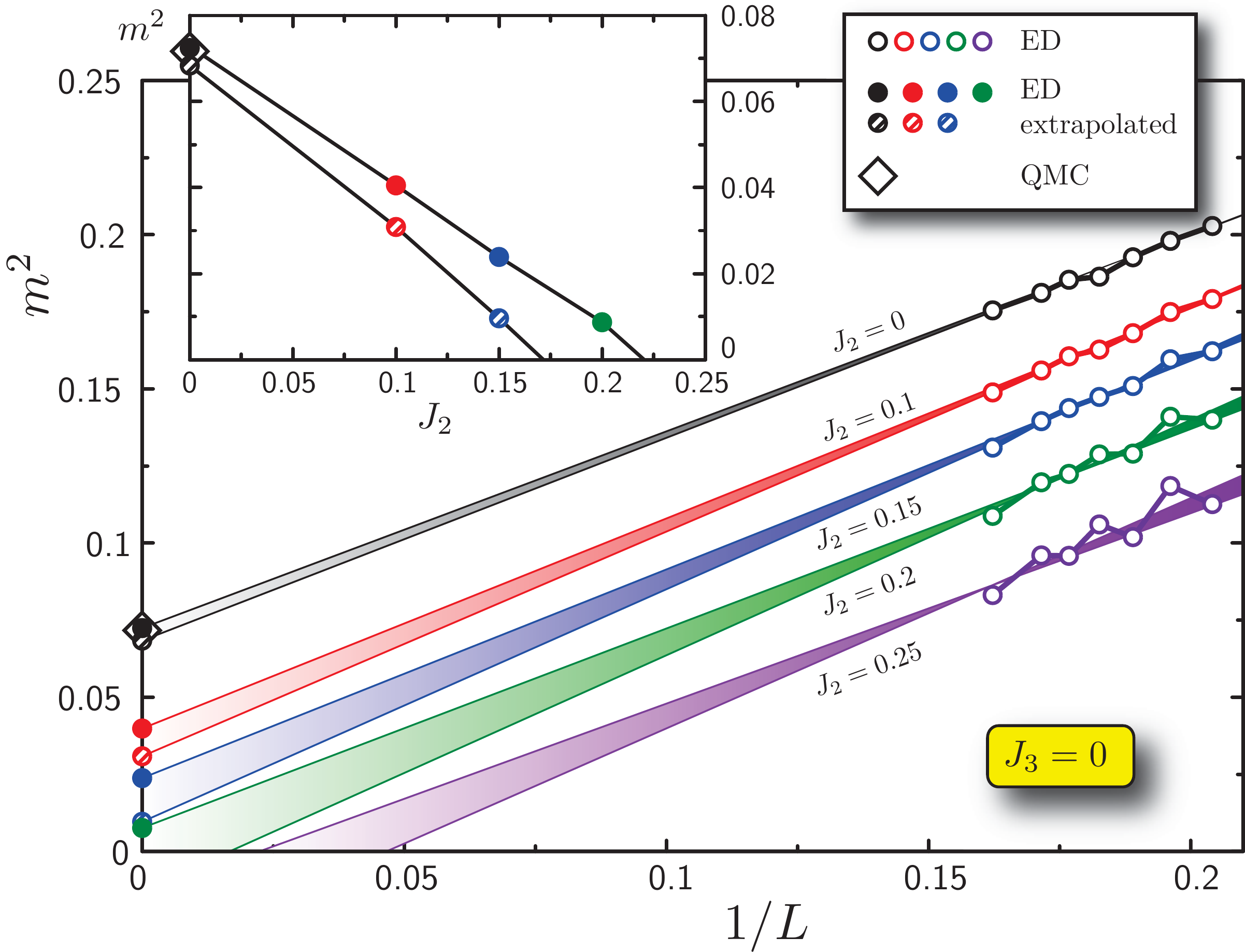}
\caption{(Color online) Squared staggered moment $m^2$ vs $1/L$ at $J_3=0$ for several $J_2$ values obtained by ED on the clusters $N=24,26,28,30,32,34,38$ and the
corresponding extrapolations to the thermodynamic limit. In the non-frustrated case $J_2=0$, we obtain a good agreement with the QMC value when using the $N=24,28$ 
and $32$ samples alone. (see text).  Inset: extrapolated value of the staggered moment $m^2_\infty$ as a function of $J_2/J_1$, vanishing between $J_2/J_1 \sim 0.17 - 0.22$,
depending on the extrapolation.
\label{fig:ED_moment}}
\end{center}
\end{figure}

The non-frustrated model ($J_2{=}J_3{=}0$) is known to possess antiferromagnetic (AF) long-range order. This has been shown by several techniques including linear spin-wave theory~\cite{Weihong1991}, a coupled cluster 
method~\cite{Krueger2000}, ED~\cite{Krueger2000,RichterLNP2004}, series expansions around the Ising limit~\cite{Oitmaa1992}, tensor network studies~\cite{Jiang2008,Xie2009,Zhao2010}, variational Monte-Carlo~\cite{Noorbakhsh2009} and quantum Monte-Carlo (QMC) simulations~\cite{Reger1989,Castro2006a,Loew2009}. In particular, the staggered moment is $m_\infty = 0.2677(6)$ \cite{Castro2006a}, a value that is significantly reduced by quantum fluctuations compared to the classical value of $1/2$. 

In Fig.~\ref{fig:ED_moment}, we plot ED data for the finite-size magnetic order parameter squared~\cite{Schulz1996}: 
\begin{equation}
m^2(N)=\frac{1}{N(N+2)} \left(\sum_i (-1)^{i}\ \mathbf{S}_i\right) ^2 
\end{equation}
for various clusters sizes $N$ and $J_2$ values (we set $J_3=0$ for the moment). Standard finite-size scaling predicts leading $1/L =1/\sqrt{N}$ 
corrections~\cite{Neuberger1989,Schulz1996}, which we find to be quite well satisfied even for small clusters for the unfrustrated case. 
The infinite system size estimate including all system sizes shown ($N=24,26,28,30,32,34,38$) is $m^2(\infty) =0.0684$, ($m_\infty=0.262$)~\footnote{An earlier ED based 
estimate~\cite{RichterLNP2004} using also up to 38 sites, but based on a different definition of the finite size $m^2(N)$ - and including also smaller systems - obtained $m=0.2788$.}.
Our best agreement with QMC is found based on the samples with $N=24,28,32$ sites only, yielding an estimate of $m^2(\infty)=0.0728$, corresponding to $m_\infty=0.270$. The
discrepancy between the different ED extrapolations results in a $\sim$ 5\% uncertainty on the magnitude of the magnetic moment. 

When $J_2$ is switched on we notice that the finite size data starts to deviate systematically from a straight line in $1/L$. We observe that systems which contain 
an $M$ point in the Brillouin zone ($N=24,28,32$), behave consistently with respect to each other - studying e.g.~the derivative $\mathrm{d} m^2/\mathrm{d}J_2$ - compared 
to the other system sizes~\footnote{This also a posteriori justifies our choice for a separate extrapolation for those systems in the unfrustrated case.}. 
We therefore choose to base one of the extrapolations (solid circles in Fig.~\ref{fig:ED_moment}) on this class of samples. The second estimate is obtained by using all
the shown system sizes (hatched circles in Fig.~\ref{fig:ED_moment}).
Now, as $J_2$ is increased starting from zero, the extrapolated staggered moment $m^2(\infty)$ decreases quite rapidly, roughly linear with increasing $J_2$ (inset of Fig.~\ref{fig:ED_moment}),
and vanishes continuously around $J_2^c = 0.17 \sim 0.22$, based on the two extrapolations. 
Despite some uncertainty, this constitutes a critical value $J_{2}^c$ which is larger than the classical estimate $1/6$~\cite{Rastelli19791}, the linear spin
wave~\cite{Fouet2001} and non-linear sigma model~\cite{Einarsson1991} results of $0.1\sim0.12$, and substantially larger than a recent variational Monte Carlo (VMC) estimate of $0.08$~\cite{Clark2010}.
Our estimate is however in agreement with a Schwinger-boson mean field treatment which reported a critical $J_2$ of about $0.2$~\cite{Mattsson1994}. 
A possible physical explanation of this shift of the transition to larger values of $J_2$ is that in some cases quantum fluctuations prefer collinear over spiral states, 
such as e.g.~in the $J_1-J_3$ model on the square lattice~\cite{Leung1996b,Mambrini2006,Reuther2011}.

We have also determined the ordered moment along a second $J_2$ cut at constant $J_3=0.3$ (data not shown). In this case the cluster size and shape dependency is even more
pronounced and makes an accurate determination of the critical $J_2$ value rather difficult. Similar extrapolations based on either all samples or only 
a subset of the samples gives a transition point somewhere between $J_2 \sim 0.27 - 0.33$, although the actual uncertainty is probably larger.

\begin{figure*}[t]
\begin{center}
\includegraphics[width=\linewidth]{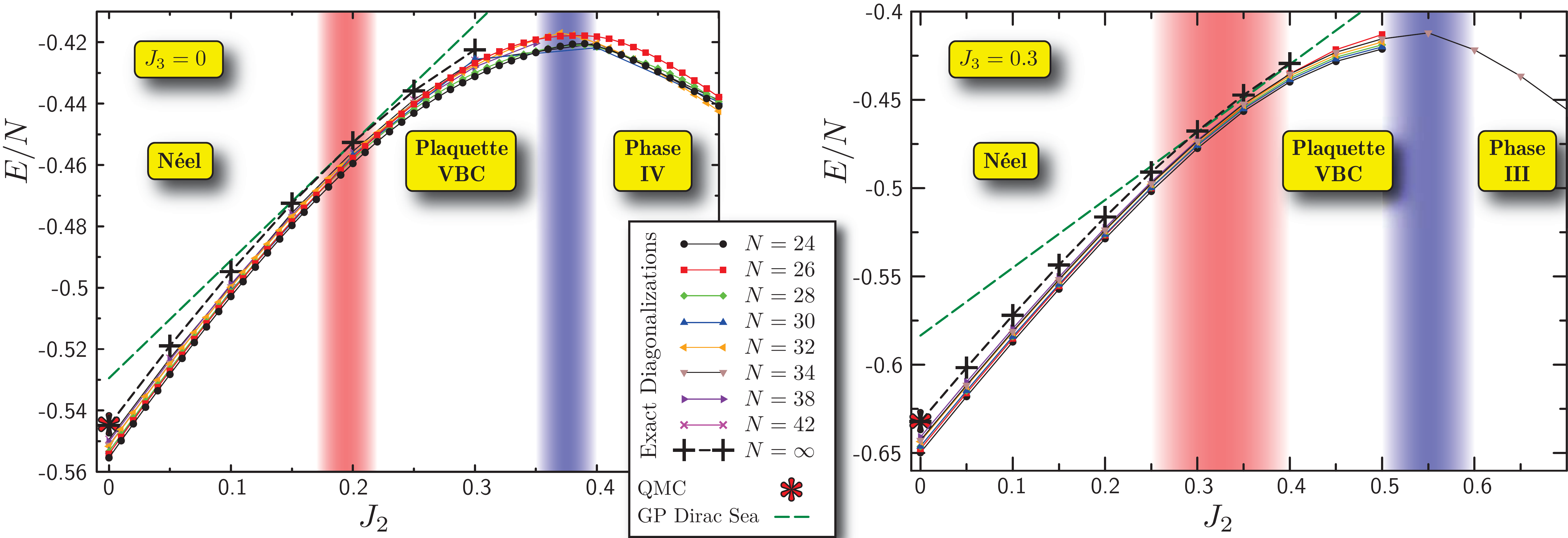}
\caption{(Color online) {\it Left Panel.} ED: Ground state energy of samples up to $N=42$ at $J_3=0$, 
together with the resulting $N\rightarrow\infty$ estimate. For comparison 
the QMC result at $J_2=0\ $\cite{Reger1989,Loew2009} and the energy of the Gutzwiller 
projected Dirac sea (see App.~\ref{sec:prop_gutzwiller}) are displayed. The light red shaded area denotes the 
approximate location of the disappearance of N\'eel order, while the dark blue shaded region denotes the 
approximate location of the first order transition to yet another phase (likely phase IV). 
{\it Right Panel.} ED: Ground state energy of samples up to $N=38$ at $J_3=0.3 J_1$,
together with the resulting $N\rightarrow\infty$ estimate. For comparison our ALPS QMC result at $J_2=0$ and the energy 
of the Gutzwiller projected Dirac sea (see App.~\ref{sec:prop_gutzwiller}) are displayed. The light red 
shaded area denotes the  approximate location of the disappearance of N\'eel order, while the dark blue
shaded region denotes the approximate location of the first order transition to yet another phase (likely phase III).
Note for both panels the good agreement between the extrapolated ED and QMC results at $J_2=0$ on the one hand, and between 
the Gutzwiller projected Dirac sea wave function and the extrapolated ED data at the magnetic to 
non-magnetic transition on the other hand.
\label{fig:gs_energy}}
\end{center}
\end{figure*}

\begin{figure}[t]
\begin{center}
\includegraphics[width=\columnwidth]{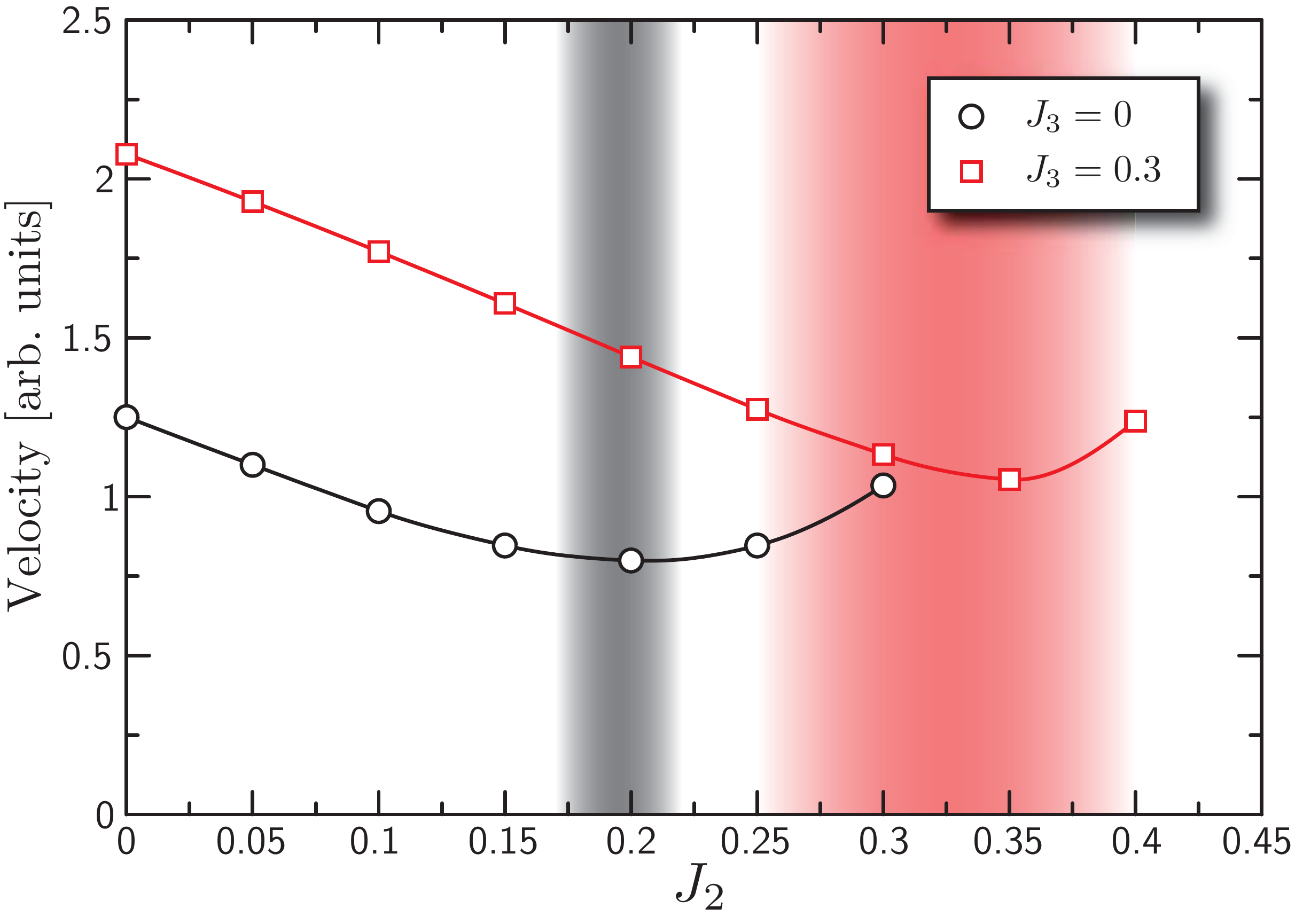}
\caption{(Color online) $J_2$ dependence of the $1/L^3$ correction term to the 
energy per site. This quantity is proportional to the spin wave velocity $c$ in the
N\'eel ordered phase. At the transition its value is expected to be finite. 
\label{fig:velocity}}
\end{center}
\end{figure}

In order to corroborate the location of the disappearance of N\'eel order, we study the energy per site along the same two constant $J_3$ lines, one located at $J_3=0$ and the other one at $J_3=0.3$.
In a N\'eel ordered phase the leading finite size corrections to the energy per site are expected to scale as
\begin{equation}
E/N=e_N=e_\infty -  \frac{\alpha\ c}{N^{3/2}} +\ldots
\label{eqn:e_correction}
\end{equation}
i.e. with a leading $1/L^3\ $ correction~\cite{Neuberger1989,Schulz1996}, and the coefficient 
of this term is proportional to the spin wave velocity $c$. In Fig.~\ref{fig:gs_energy} ($J_3=0$ on the left panel and $J_3=0.3$ on the right panel) we display the energy per site of samples of
up to 42 spins in the relevant $J_2$ range together with the resulting $N\rightarrow\infty$ estimate $e_\infty$. We extrapolate the energy according to Eq.~\eqref{eqn:e_correction} up to $J_2 =0.3$ for
$J_3=0$, and up to $J_2=0.4$ for $J_3=0.3$.  As can be seen in Fig.~\ref{fig:velocity}, the prefactor of the $1/N^{3/2}$ correction term is reduced upon approaching the transition region, but seems to
stay constant at the transition, in analogy to the frustrated square lattice antiferromagnet~\cite{Schulz1996,Richter2010b}. Note that the shaded regions denote the approximate locations of the transitions
based on the extrapolation of the ordered moment, and the minimum of the velocity agrees reasonably well with those estimates.

Returning to the extrapolated energies, we note that for the unfrustrated $J_2,J_3=0$ case in Fig.~\ref{fig:gs_energy} (left panel) the extrapolated energy per site $e_\infty$ is in very good agreement with 
published QMC results~\cite{Reger1989,Loew2009}, and we find similarly good agreement for $J_2=0, J_3=0.3$ in Fig.~\ref{fig:gs_energy} (right panel), where we performed ALPS SSE 
simulations~\cite{alet:05a,albuquerque:07} to obtain an accurate estimate for the energy.
For both $J_3$ values the energy then first rises almost linearly with increasing $J_2$, as expected for this particular N\'eel phase (note that the derivative $\mathrm{d}e/\mathrm{d}J_2$ is proportional to
$\langle \mathbf{S}_i\cdot\mathbf{S}_j\rangle$ on the $J_2$ bonds as a consequence of the Hellmann-Feynman theorem). The energy curves flatten at larger $J_2$ and exhibit a maximum
around $J_2\sim 0.35-0.4$ for $J_3=0$ and $J_2\sim0.5-0.6$ for $J_3=0.3$. A comparison with the fidelity data shown in Fig.~\ref{fig:phase_diagram}, suggests that the maximum of the energy approximately coincides
 with the avoided level crossing to a different phase.

\begin{figure*}[t]
\begin{center}
\includegraphics*[width=0.75\linewidth]{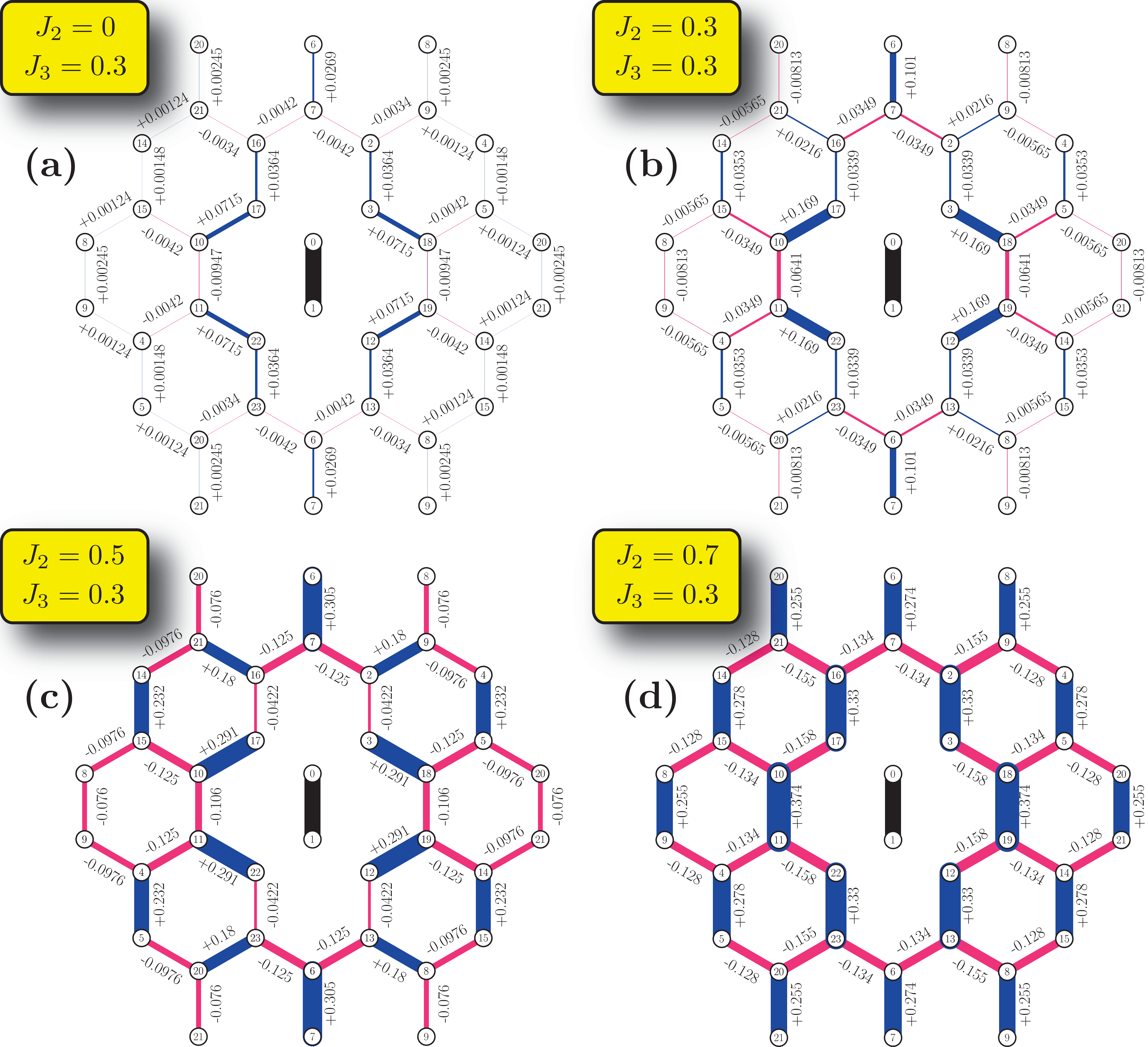}
\caption{(Color online) Four-spin correlations [Eq.~(\ref{eq:fourpoint})] on an $N=24$ site cluster for several different $J_2$ values and constant $J_3=0.3$:
		{\bf (a)} $(0.0,0.3)$ in the N\'eel phase I, {\bf (b)} $(0.3,0.3)$ in region of the phase transition from phase I to V, {\bf (c)} $(0.5,0.3)$
		in the {\em plaquette} phase V, and {\bf (d)} $(0.7,0.3)$ in the lattice nematic ({\em staggered} dimer VBC) phase III or IV.
		The reference bond is indicated by the thick-black line. Negative (positive) correlations are represented by red (blue) bonds.
		\label{fig:FourPoint_N24_many}}
\end{center}
\end{figure*}

Inspired by the success of a simple Gutzwiller projected half-filled tight-binding wave function on the triangular lattice in describing the spin liquid regime on the insulating side of the Mott transition~\cite{Motrunich2005},
we have analyzed a related wave function on the honeycomb lattice: the Gutzwiller projected half-filled honeycomb tight binding wave function (termed the Gutzwiller projected "Dirac sea" in the following). 
This wave function is discussed in some detail in App.~\ref{sec:prop_gutzwiller}. It is a \emph{parameter-free} 
variational wave function, and its energy for the Hamiltonian considered here is given in 
Eq.~\eqref{eqn:dirac_energy}. This energy is plotted using a dashed green line in Fig.~\ref{fig:gs_energy}. Quite remarkably the energy is very close to the finite size extrapolated 
ED energies precisely in the region where the N\'eel order is about to vanish (light red shaded uncertainty regions). As discussed in App.~\ref{sec:prop_gutzwiller}, the Gutzwiller projected Dirac sea wave function has 
algebraically decaying spin-spin correlations with a sign structure that is identical to the one displayed by the N\'eel state. With its algebraically decaying correlation this wave function could
in principle describe qualitatively a putative continuous quantum phase transition from the N\'eel ordered phase to a quantum paramagnet. Inspecting the nature of the dimer-dimer correlations in the Gutzwiller projected Dirac sea, the signs of the dimer-dimer correlations are identical to the ones in the {\em columnar} (Read-Sachdev) or {\em plaquette} VBC (disregarding one particular distance highlighted in Fig.~\ref{fig:GPDS_dimer_correlations}).
Given the surprisingly accurate energy of this wave function at the transition, a plausible scenario is the presence of a continuous N\'eel to {\em columnar/plaquette} VBC quantum phase transition in this frustrated honeycomb
antiferromagnet. We will discuss this scenario and other possibilities in more detail later on. Let us now consider the dimer-dimer correlations in the magnetically disordered phase, in order to verify whether there is indeed a VBC phase present.

\begin{figure*}[!ht]
\begin{center}
\includegraphics[width=0.8\linewidth]{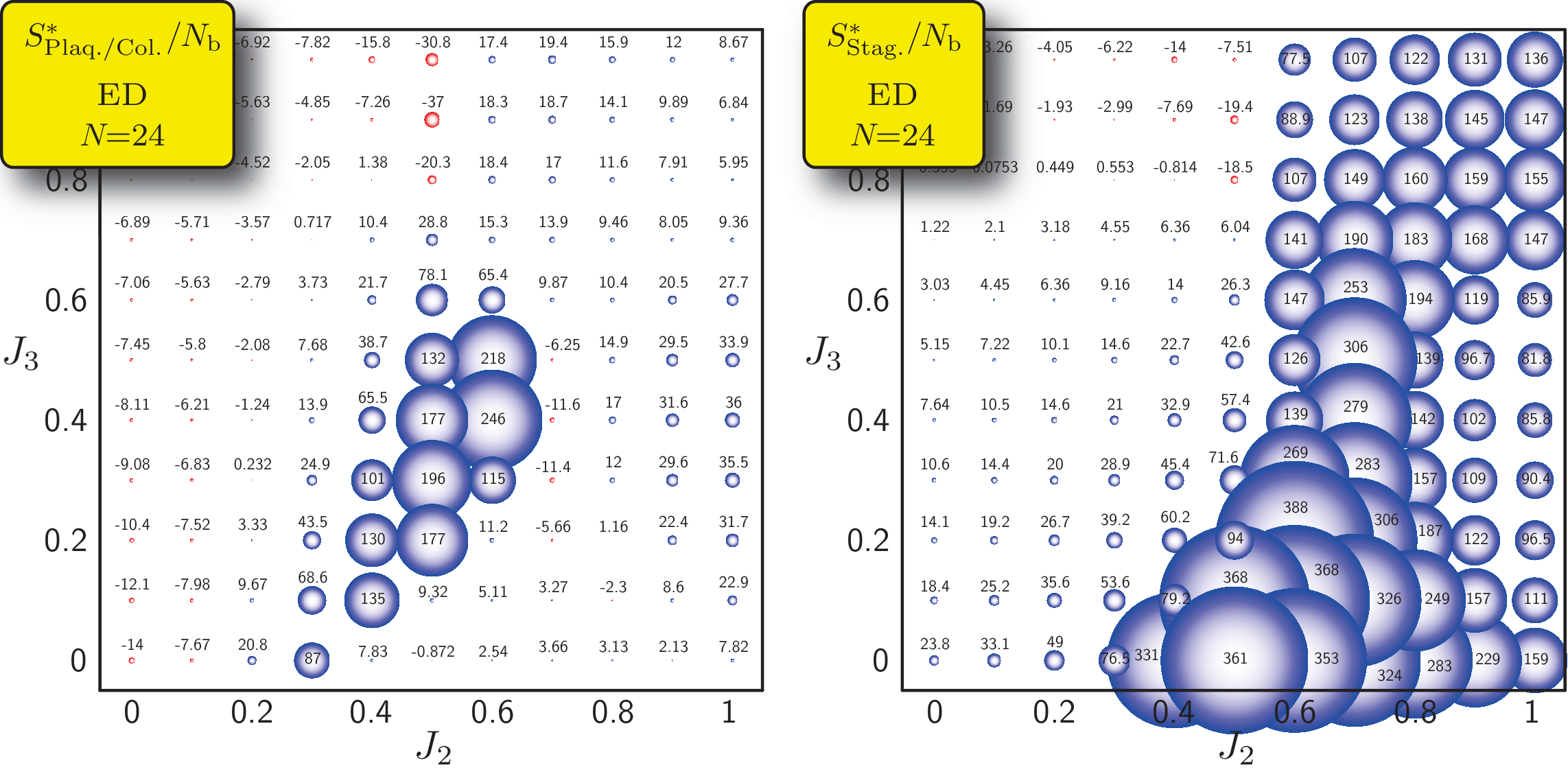}
\caption{(Color online) Left panel: {\em plaquette/columnar} VBC structure factor $S_{\rm Plaq./Col.}^{\ast}/N_{\rm b}$ for $N=24$, obtained using ED in the $S^z$ basis, as a function of
                                          $(J_2,J_3)$. The radius of the circles is proportional to $S_{\rm Plaq./Col.}^{\ast}/N_{\rm b}$. Numbers correspond to $10^3 S_{\rm Plaq./Col.}^{\ast}/N_{\rm b}$.
                                     Right panel: {\em staggered} VBC structure factor $S_{\rm Stag.}^{\ast}/N_{\rm b}$ for $N=24$, obtained using ED in the $S^z$ basis, as a function of
                                          $(J_2,J_3)$. The radius of the circles is proportional to $S_{\rm Stag.}^{\ast}/N_{\rm b}$. Numbers correspond to $10^3 S_{\rm Stag.}^{\ast}/N_{\rm b}$.
                                          Note that the strong {\em plaquette/columnar} signal is found within phase (V) of Fig.~\ref{fig:phase_diagram}, while the strong {\em staggered} signal is associated to the phases (II),(III) and (IV).
\label{fig:NNVB_dimer_structure_factors_ed}}
\end{center}
\end{figure*}

\subsection{Dimer correlations}
\label{subsec:ed_dimercorrs}

{\it Real space correlations.} In this section we study dimer-dimer correlations using ED. Our aim is to highlight the structure of 
the correlations along a $J_2$-cut at constant $J_3=0.3$ for the $N=24$ sample~\footnote{The next larger
sample which is compatible with the expected VBCs and has sixfold rotation symmetry consists of 42 spins. While we
have been able to obtain energies for this systems, calculating four-spin expectation values along a whole cut 
was prohibitive with our given resources.}. 
We measure the following four-spin correlation function:
\begin{equation}
C_{ijkl}= 4 \left( \langle \left({\mathbf S}_{i} \cdot {\mathbf S}_{j} \right)\left({\mathbf S}_{k} \cdot {\mathbf S}_{l} \right) \rangle - 
                \left( \langle {\mathbf S}_{i} \cdot {\mathbf S}_{j} \rangle \right)^{2} \right)~,
 \label{eq:fourpoint}
\end{equation}
where $i,j$ and $k,l$ are nearest-neighbor bonds on the honeycomb lattice. In Fig.~\ref{fig:FourPoint_N24_many} correlation
function results for four different values of $J_2$ are shown. The panel (a) shows the dimer-dimer correlations deep in the N\'eel phase
at $J_2=0, J_3=0.3$, where we expect the correlations to decay rapidly, but with a power-law, due to the coupling to the multi spin
wave continuum. In panel (b), at $J_2=0.3, J_3=0.3$, we sit approximately at the N\'eel to paramagnet transition, and 
some of the more distant bonds have changed sign compared to the N\'eel phase. Note that this correlation pattern matches qualitatively
the one of the Gutzwiller projected Dirac sea discussed in App.~\ref{sec:prop_gutzwiller} (as well as the $\mathbb{Z}_2$-liquid discussed 
in Ref.~\onlinecite{Clark2010}), and surprisingly also the one reported for the spin liquid regime of the half-filled Hubbard model in Ref.~\onlinecite{Meng2010}. 
Panel (c) at $J_2=0.5, J_3=0.3$ shows pronounced and long-ranged correlations, which at first sight seem to be compatible with either a {\em columnar} (Read-Sachdev) or 
{\em plaquette} VBC according to App.~\ref{sec:PureStates}. A more quantitative inspection reveals however that the largest distance positive/negative
correlations are respectively close to 0.18 and -0.0935, which is in favor of a ($d$-wave) {\em plaquette} phase. We revisit the question of {\em columnar} versus
{\em plaquette} order again in the context of the effective quantum dimer model, and corroborate the present finding of a {\em plaquette} phase. Finally, panel (d) at
$J_2=0.7, J_3=0.3$ shows very strong correlations reminiscent of a {\em staggered} dimer phase. We stress again that this finding alone does not
discriminate between a spin gapped valence bond crystal or a magnetically ordered phase of type (II), (III) or (IV). While a {\em staggered} valence bond
crystal neighboring the {\em plaquette} phase is likely (according to Refs.~\onlinecite{Fouet2001,Mulder2010,Mosadeq2010}), at larger $J_2$ the {\em staggered} signal
in the dimer-dimer correlations could persist despite the appearance of magnetic order. 

\begin{figure}[!ht]
\includegraphics[width=0.95\linewidth]{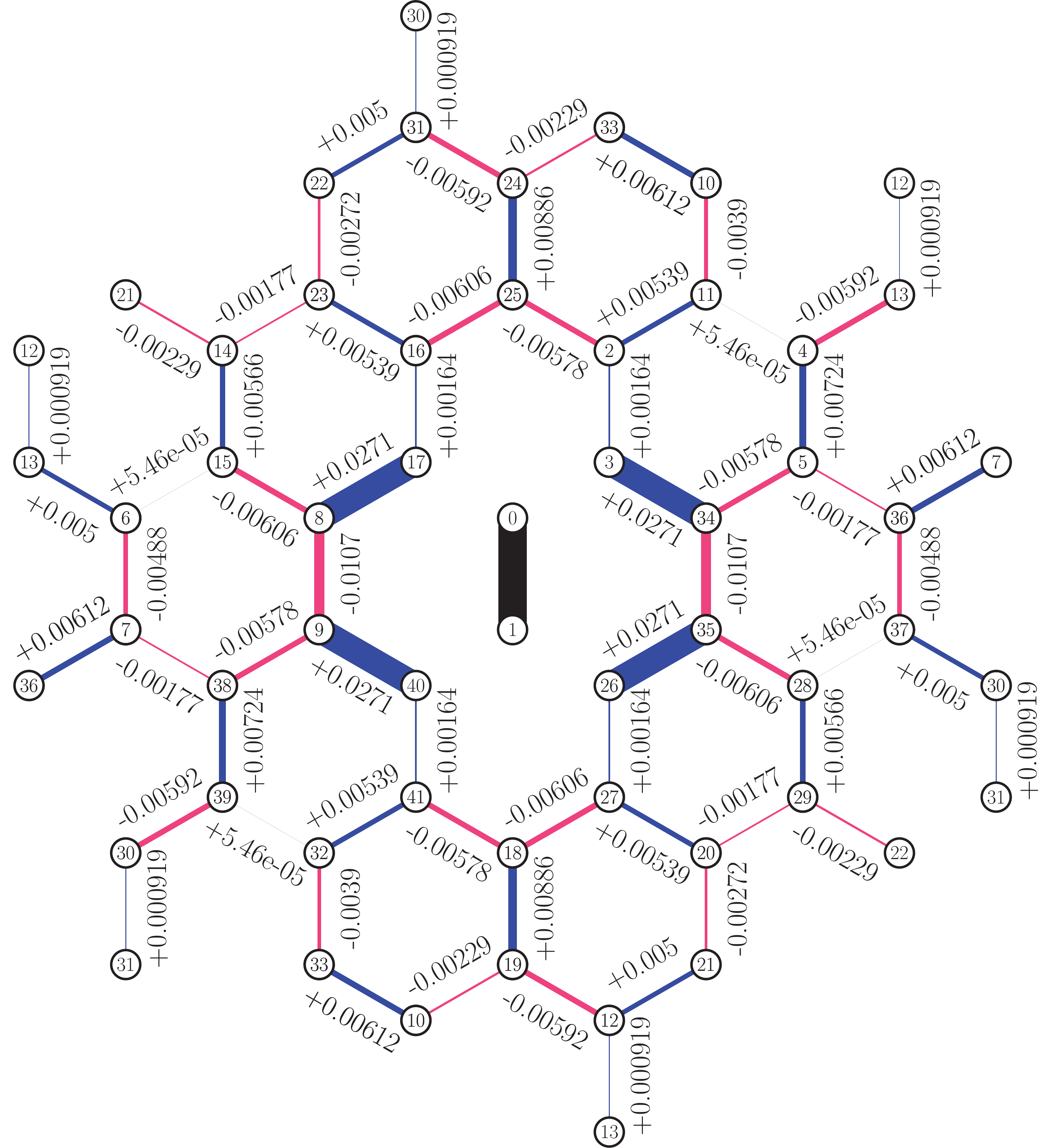}
\caption{(Color online) $S^z$ version of the dimer correlations obtained using ED in the $S^z$ basis for $N=42$ at $J_2=0.3, J_3=0$.
\label{fig:dimer_N_42_ED}}
\end{figure}

{\it VBC structure factors.} It is instructive to grasp correlations using integrated quantities such as dimer structure factors.
As displayed in Fig. \ref{fig:CorrelationSnapshot} of App.~\ref{sec:PureStates}, two different dimer correlation patterns emerge for  {\em plaquette/columnar} states or a  {\em staggered} state. In order to detect them, we define the dimer structure factors as:
\begin{equation}
S_\alpha = \sum_{\langle k,l \rangle} \varepsilon_\alpha (k,l) C_{ijkl}~,
 \label{eq:structure}
\end{equation}
with $\alpha={\rm Plaq./Col.}$ or $\alpha = {\rm Stag.}$, where $\varepsilon_\alpha (k,l) = +1$ if ${\langle k,l \rangle}$ are NN sites such that $C_{ijkl} \geq 0$ for ``pure" $\alpha$ states and $\varepsilon_{\rm VBC}(k,l) = -2$ otherwise (strong correlations the closest to the reference bond are not included for the related quantity $S_\alpha^{\ast}$; see Ref.~\onlinecite{Mambrini2006}). It is important to stress here that the $S_\alpha$
are order parameters detecting lattice symmetry breaking, which do not distinguish themselves between genuine VBC ordering or a lattice symmetry breaking magnetic state.

A full scan of dimer structure factors associated to either {\em plaquette/columnar} or {\em staggered} valence bond crystal order is shown in Fig.~\ref{fig:NNVB_dimer_structure_factors_ed}. Consistently with real space dimer correlation analysis, two phases come up with strong {\em plaquette/columnar} (left panel) or {\em staggered} (right panel) signal. The {\em staggered} signal is especially strong in the vicinity of the $J_3=0$ line and close to the avoided level crossing. In contrast, the {\em columnar/plaquette} signal is strongest around $J_2 \sim 0.6, J_3\sim 0.4$, and decreases upon approaching the $J_3=0$ line. In order to address the
behavior at $J_3=0$ we have also calculated the $S^z$ dimer correlations [cf. Eqn.~(\ref{eqn:sz_dimer_corrs})] for $J_2=0.3, J_3=0$ on the $N=42$ sample, which would be compatible with a {\em columnar/plaquette} VBC. The corresponding plot shown in Fig.~\ref{fig:dimer_N_42_ED} exhibits a correlation pattern reminiscent of the one expected for {\em columnar/plaquette} states, but the correlations 
are not particularly strong, and also exhibit a few defects in the form of bond which show inverted correlations compared to the {\em columnar/plaquette} expectations. We are thus currently unable to 
discriminate whether this picture corresponds to a {\em columnar/plaquette} VBC with a small order parameter or a genuine spin liquid, and more work is needed to clarify the behavior at $J_3 \sim 0$.

\section{Exact Diagonalizations in the Valence-Bond Basis and Quantum Dimer Models}
\label{sec:EDVB}

\subsection{Diagonalization in the Nearest-Neighbor Valence-Bond Basis}

In this subsection we present results obtained from exact diagonalizations in the (variational) basis given by the set of nearest-neighbor
valence-bond states~\cite{mambrini00,Mambrini2006}. Recently Mosadeq {\em et al.}~\cite{Mosadeq2010} presented an analysis using the same
technique but limited to $J_3=0$ and small system sizes ($N=54$). Here we consider the more general case of finite $J_3 \in [0,1]$
and investigate considerably larger clusters ($N=72$ for correlations and $N=96$ for energies), allowing us to perform systematic finite size extrapolations. 

\subsubsection{The Method}
\label{subsubsec:NNVB}

When frustration is dominant and destabilizes magnetic phases it is possible to explicitly take into account that states with non-zero total spin are
unimportant in accounting for the low-energy physics and to describe the system solely in terms of the $S=0$ subspace. This subspace
can be spanned by the set of arbitrarily ranged valence-bond (VB) states~\cite{Hulten38,Karbach93}, which forms an over-complete basis and
is thus difficult to manipulate, especially in numerical studies. A natural way of circumventing this difficulty is to impose a cutoff on the maximum
range of the VBs to be considered; in particular, it is possible to devise an approach where only nearest-neighbor VB (NNVB) states are taken into
account.\cite{mambrini00,Mambrini2006} While the restriction to NNVB states is obviously a variational approximation, it offers the key numerical advantage
of a significant reduction of the Hilbert space and has been shown to yield sound results for a number of strongly frustrated models, whose
low-energy physics is dominated by short-range spin-singlets.\cite{mambrini00,fouet03,Mambrini2006}

We briefly recall how the method can be applied and refer to Refs.~\onlinecite{mambrini00,Mambrini2006} for details. We follow a heuristic argument
and try to formulate the eigenvalue problem in the restricted NNVB subspace $\{ \ket{\varphi_i} \}$ simply as:
$\sum_i \alpha_i \hat{{\cal H}} \ket{\varphi_i} = E \sum_i \alpha_i  \ket{\varphi_i}$. However, since the set $\{ \ket{\varphi_i} \}$ is not invariant under 
the application of ${\cal H}$, this relation cannot hold in the particular singlet subspace but can explicitly be enforced in the restricted NNVB subspace by considering
\begin{equation}
\sum_i \alpha_i \bra{\varphi_j} \hat{{\cal H}} \ket{\varphi_i} = E \sum_i \alpha_i \braket{\varphi_j}{\varphi_i}~,
 \label{eq:GEP}
\end{equation}
for all $\ket{\varphi_j} \in \{ \ket{\varphi_i} \}$. This last equation is nothing but a generalized eigenvalue problem (GEP) for the two matrices
with elements given by ${\cal H}_{ij}=\bra{\varphi_j} \hat{\cal H} \ket{\varphi_i}$ and ${\cal O}_{ij} = \braket{\varphi_j}{\varphi_i}$, the latter
explicitly denoting the non-orthogonality of NNVB states. However, it is crucial here that despite of their non-orthogonality, the NNVB states are 
linearly independent on most relevant lattices~\cite{Seidel09,Wildeboer11}, and particularly on the honeycomb lattice with periodic boundary
conditions considered here~\cite{Wildeboer11}.
GEPs are computationally more demanding than conventional eigenvalue problems,
especially in the present case where both ${\cal H}_{ij}$ and ${\cal O}_{ij}$ are dense matrices. In spite of this, since for a given
system size the dimension of the NNVB subspace is much smaller than that of the total $S^{z}=0$ subspace, the method discussed here allows 
us to treat considerably larger clusters, and thus to perform more extended finite size extrapolations,
than possible within conventional ED. We also remark that, as for standard ED, it is possible to take advantage of lattice symmetries, so that the 
size of the matrices to be considered is further reduced and, even more importantly, crucial information on quantum numbers is readily available.

\begin{figure}
  \begin{center}
    \includegraphics*[width=\linewidth]{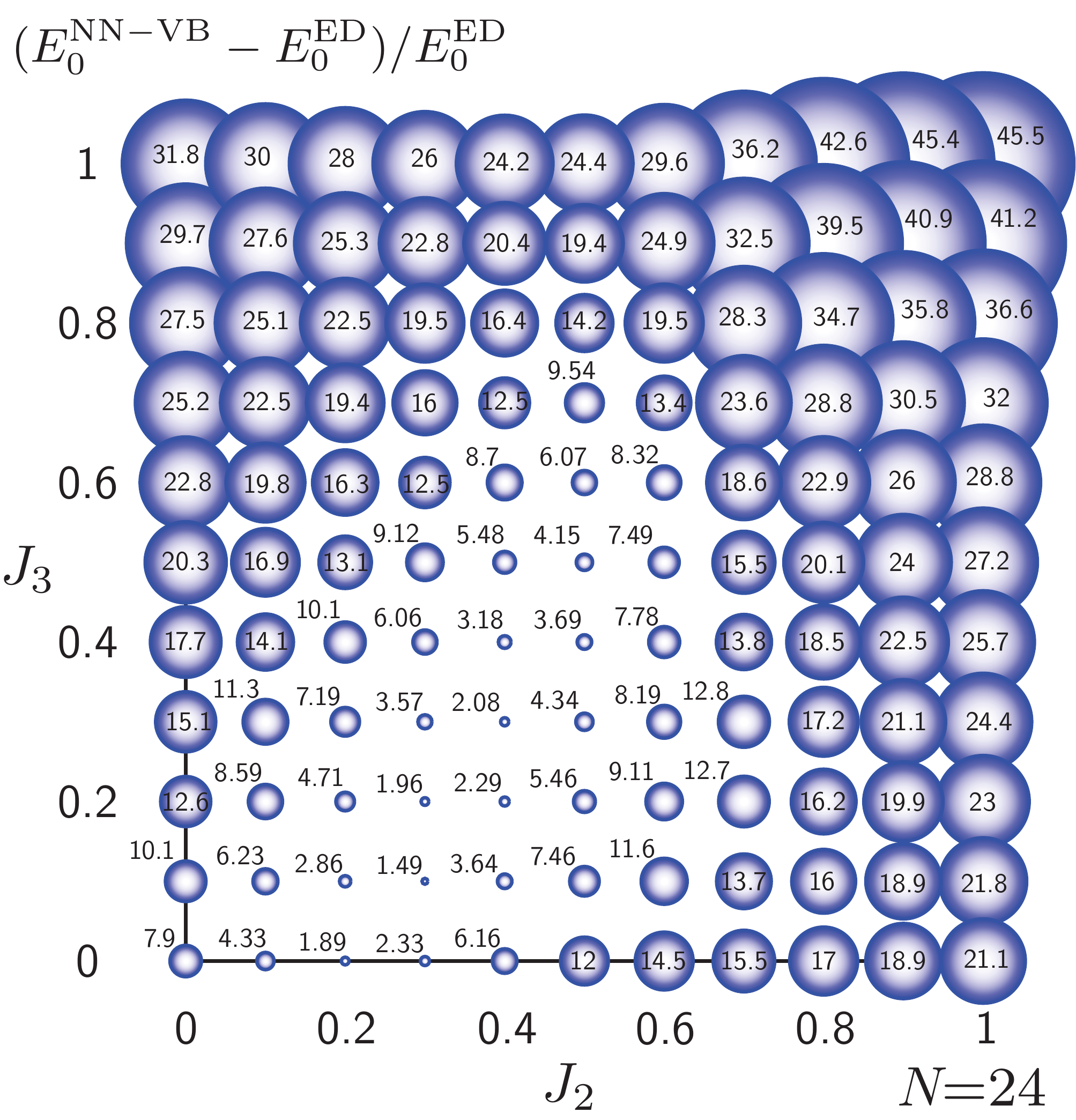}
   \end{center}
   \caption{(Color online) Comparison between the ground-state energy obtained from diagonalization in the NNVB and in the $S^{z}$ basis, as a function
                   of $J_2, J_3 \in [0, 1]$. The radius of the circles is proportional to the percent relative error, $(E_{0}^{\rm NNVB} - E_{0}^{\rm ED}) / E_{0}^{\rm ED} \times 100$.
                   Data obtained from a cluster comprising $N=24$ sites.}
  \label{fig:E0_Error}
\end{figure}

Despite such appealing features, due to its variational nature the method just described lacks built-in indicators of its own reliability. This drawback
can be circumvented by relying on unbiased techniques, such as ED in the $S^z$ basis, that are used in providing benchmarks to validate the 
restriction to the NNVB manifold. As a first step in this direction, in Fig.~\ref{fig:E0_Error} we plot results for the relative difference between the ground-state energy
of the model Eq.~(\ref{eqn:HeisenbergHamiltonian}) obtained by solving the GEP in the NNVB basis and from ED in the $S^{z}$ basis,
$(E_{0}^{\rm NNVB} - E_{0}^{\rm ED}) / E_{0}^{\rm ED}$, for an $N=24$ site cluster. (Qualitatively similar results are obtained for a less symmetric
cluster with $N=30$ sites, not shown here. We note however that finite size effects in the NNVB energy per site seem to be surprisingly large, as shown in
App.~\ref{sec:comp_NNVB_QDM}. The qualitative result regarding the region of best match with $S^z$ ED seems to be stable with system size however.)  
As expected, since long-range VBs are required in accounting for long-range spin correlations on
two-dimensional lattices~\cite{liang88}, $E_{0}^{\rm NNVB}$ compares poorly against $E_{0}^{\rm ED}$ for couplings expected to support magnetic
phases (see sections~\ref{sec:PhDiag}, \ref{sec:SCMFT}, and \ref{sec:ED}). Conversely, $(E_{0}^{\rm NNVB} - E_{0}^{\rm ED}) / E_{0}^{\rm ED} \lesssim
5\%$, in the region of the phase diagram where magnetically-disordered phases are likely to be stabilized, cf.~Fig.~\ref{fig:phase_diagram}; in particular, for 
$J_3=0$ small relative errors are observed for $0.2 \lesssim J_2 \lesssim 0.3$, in agreement with Ref.~\onlinecite{Mosadeq2010}. 
The fact that a small number of VB configurations ({\em eight} NNVB states, as opposed to $19873$ in ED in the $S^{z}=0$ subspace - lattice symmetries being
exploited in both cases) is able to reproduce the GS energy in an extended region of the parameter space up to a relative
error that can be as small as $\sim 1.5\%$ constitutes good evidence that the NNVB subspace may
be able to capture the low energy physics of the magnetically-disordered phases. Of course, to confirm this statement it is crucial to go beyond a
simple energy-based criterion and to compare the nature of the correlations contained in this variational wave function and the exact one. In what follows
we proceed to a thorough characterization of four-spin correlations.
 
\subsubsection{Four-Spin Correlations}
\label{sec:fourpoint}

\begin{figure}[!htb]
\begin{center}
\includegraphics*[width=0.65\linewidth]{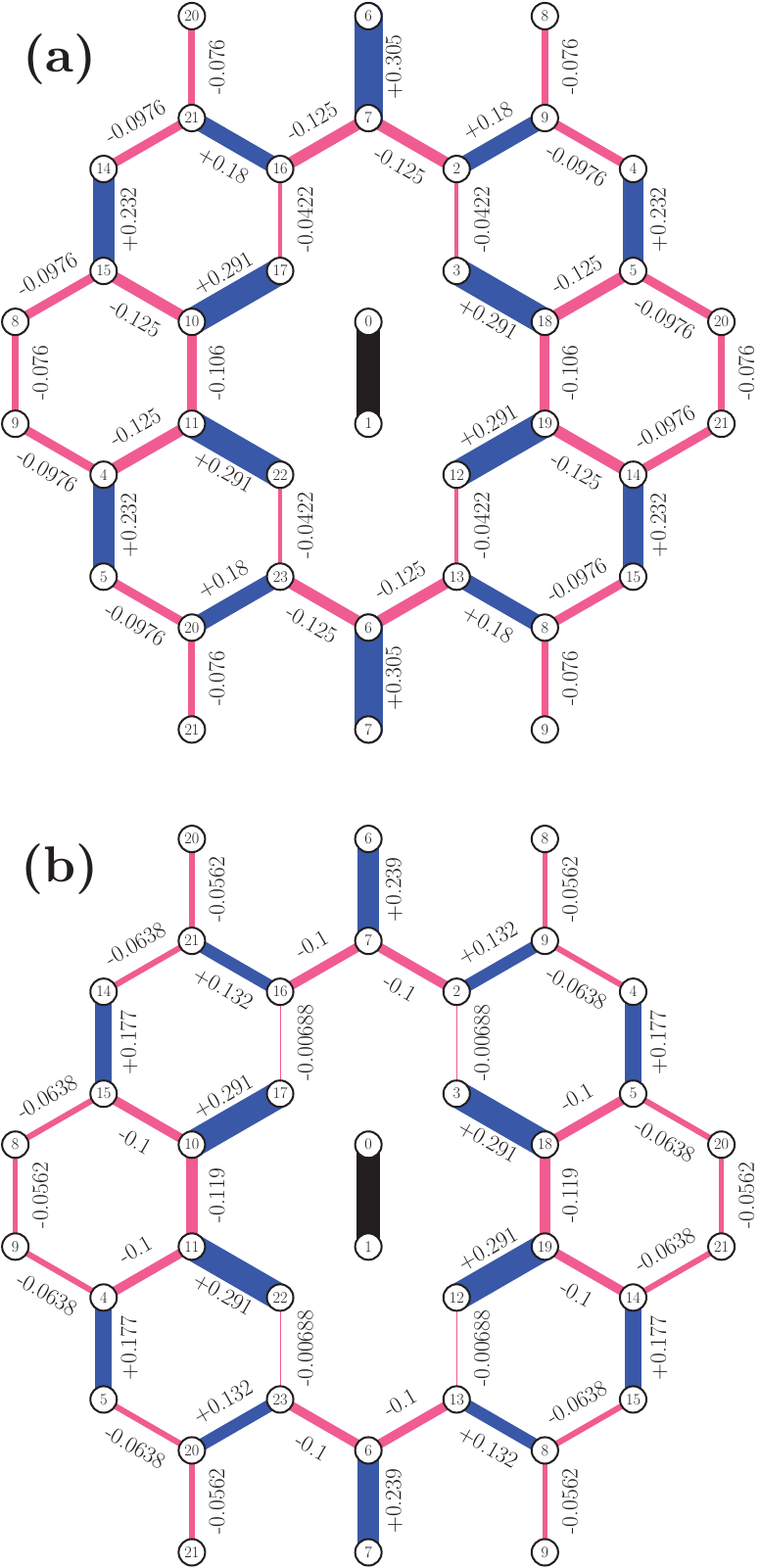}
\caption{(Color online) Four-spin correlations [Eq.~(\ref{eq:fourpoint})] on an $N=24$ site cluster for $(J_2,J_3)=(0.5,0.3)$, obtained
                from (a) ED in the $S^z$ basis and (b) by solving the GEP [Eq.~(\ref{eq:GEP})] in the NNVB subspace. The reference bond, in both panels, is
                indicated by the thick-black line.   
\label{fig:FourPoint_N24}}
\end{center}
\end{figure}

{\it Real space correlations.} We compute the four-spin connected correlation function $C_{ijkl}$ as defined in Eq.~\eqref{eq:fourpoint} in the ED section,
where $i,j$ and $k,l$ are pairs of NN sites (dimers) on the honeycomb lattice. $C_{ijkl}$ is readily evaluated by analyzing the loop structure in the
transition graphs $\braket{\varphi_j} {\varphi_i}$ for non-orthogonal NNVB states, in terms of which the lowest-energy state, solving the GEP,
[Eq.~(\ref{eq:GEP})] is expressed (for technical details on how to compute expectation values for NNVB states see Ref.~\onlinecite{beach06}).

We once more gauge the validity of our variational approach and compare the so-obtained results for $C_{ijkl}$ against those from ED for the
$N=24$ site cluster and $(J_2,J_3)=(0.5,0.3)$ in Fig.~\ref{fig:FourPoint_N24}. Semi-quantitative agreement is observed, and interestingly the correlations
obtained from the variational approach seem to be systematically smaller than those from ED, suggesting that the exclusion of longer-range
VBs has the effect that VBC order is {\em underestimated} (see below). Regarding the particular kind of VBC order that is stabilized, the two
sets of data in Fig.~\ref{fig:FourPoint_N24} are consistent with both {\em columnar} (Read-Sachdev) and {\em plaquette} VBC order, although the particularly strong
correlations at the shortest range (those involving the dimers closest to the reference bond) seemingly favor the later scenario (see the Appendix
\ref{sec:PureStates}), as vindicated in Refs.~\onlinecite{Fouet2001,Mosadeq2010}. We remark that evidence in favor of {\em plaquette} VBC
order is also found from the histogram analysis in the framework of the effective QDM presented in Sec.~\ref{sec:EDQDM}.

\begin{figure}[!htb]
\begin{center}
\includegraphics*[width=0.95\linewidth]{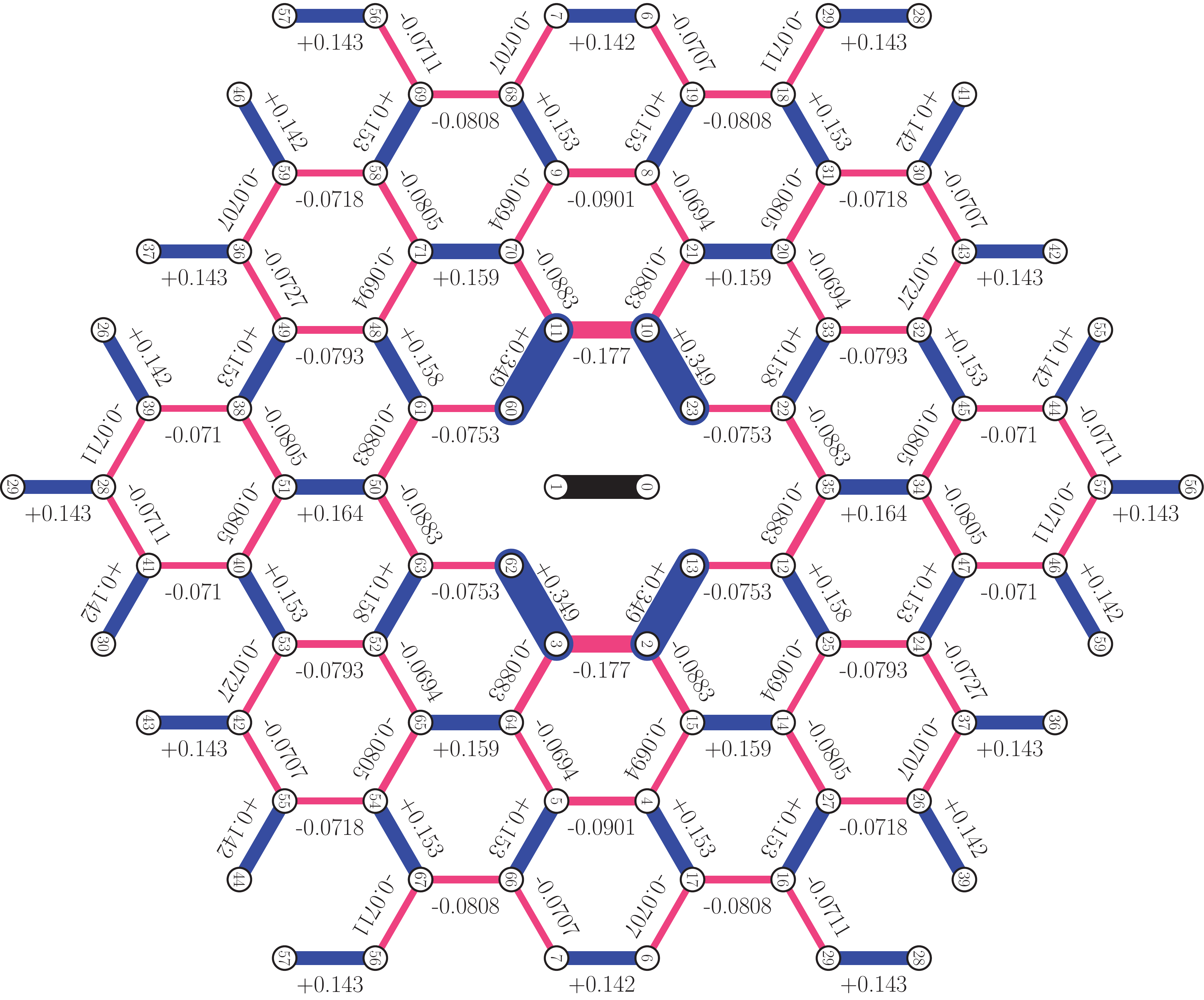}
\caption{(Color online) Four-spin correlations [Eq.~(\ref{eq:fourpoint})] on an $N=72$ site cluster for $(J_2,J_3)=(0.5,0.3)$, obtained by solving the
                GEP Eq.~(\ref{eq:GEP}) in the NNVB subspace. The thickness of the bonds is proportional to $C_{ijkl}$ and dark-blue (pale-red) indicates
                positive (negative) values. The reference bond is indicated by the thick-black line.   
\label{fig:FourPoint_N72}}
\end{center}
\end{figure}

\begin{figure}[!htb]
\begin{center}
\includegraphics* [width=0.5\textwidth]{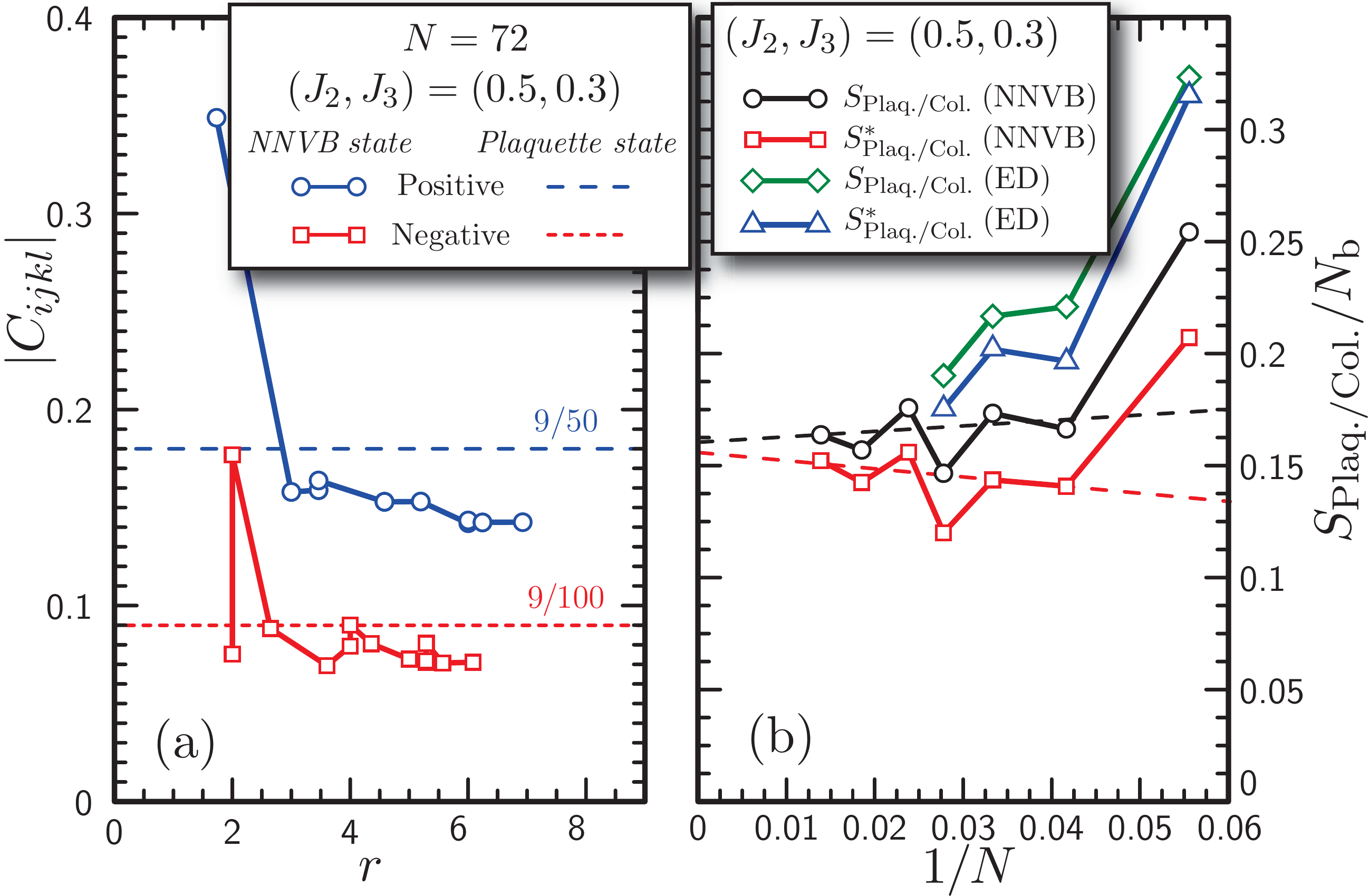}
\caption{(Color online) (a) $|C_{ijkl}|$ [Eq.~(\ref{eq:fourpoint})] as a function of distance $r$ for the $N=72$ site cluster and $(J_2,J_3)=(0.5,0.3)$, obtained
                by solving the GEP Eq.~(\ref{eq:GEP}) in the NNVB subspace; positive and negative correlations are discriminated. (b) VBC structure factor per
                per number of bonds, $S_{\rm Plaq./Col.} / N_{\rm b}$, as a function of the inverse system size $N^{-1}$ for $(J_2,J_3)=(0.5,0.3)$, as obtained from ED
                and from the variational approach in the NNVB basis. $S_{\rm Plaq./Col.}^{\ast}$ denotes the structure factor obtained by eliminating the strongest
                correlations encircling the reference bond and linear fits (full lines; data for $N=18$ and $N=36$ are excluded from the fit) extrapolate the data
                to the thermodynamic limit. In both panels, dotted lines are only guides to the eye.
\label{fig:CorrDec}}
\end{center}
\end{figure}

We take advantage of the substantially reduced dimension of the NNVB subspace and compute four-spin correlations for considerably
larger clusters, comprising up to $N=72$ sites. The spatial dependence of $C_{ijkl}$ is depicted in Fig.~\ref{fig:FourPoint_N72} for the
$N=72$ cluster and $(J_2,J_3)=(0.5,0.3)$. An even stronger resemblance to the correlation pattern for ``pure" {\em columnar} and {\em plaquette}
states is observed than for the smaller cluster with $N=24$ sites [Fig.~\ref{fig:FourPoint_N24}(b)], suggesting that the observed VBC
pattern is not merely a finite size effect. This observation is further corroborated by the data in Fig.~\ref{fig:CorrDec}(a), where
we plot $|C_{ijkl}|$ as function of $r$ (the distance between dimers $i,j$ and $k,l$) also for $N=72$ and $(J_2,J_3)=(0.5,0.3)$: $|C_{ijkl}(r)|$
decays slowly with $r$, consistent with saturation at large distances, and furthermore positive correlations are approximately twice as
large as negative ones, as expected for pure {\em columnar} (Read-Sachdev) and {\em plaquette} VBC states.


{\it Dimer structure factors.} Reliable extrapolation to the thermodynamic limit is achieved from the analysis of VBC structure factors $S_\alpha$ (see Eq.~\ref{eq:structure}). $S_\alpha$ is expected to scale like $C^{\infty}_\alpha + A/N$ and thus the existence
of $\alpha$ VBC phases is signaled by a finite value of the bond order parameter $C^{\infty}_\alpha$. In Fig.~\ref{fig:CorrDec}(b) we plot $S_{\rm Plaq./Col.}/N_{\rm b}$
and $S_{\rm Plaq./Col.}^{\ast}/N_{\rm b}$ ($N_{\rm b}$ is the total number of bonds considered in the sum) as a function of inverse system size, as obtained from both 
$S^z$-ED and by
solving Eq.~(\ref{eq:GEP}), for $(J_2,J_3)=(0.5,0.3)$. We first notice that, in agreement with our previous remark in connection to Fig.~\ref{fig:FourPoint_N24},
larger values for $S_{\rm Plaq./Col.}$ and $S_{\rm Plaq./Col.}^{\ast}$ are obtained with $S^z$-ED and that the restriction to the NNVB manifold seemingly
underestimates VBC order in the present case. Linear fits to the NNVB data shown in Fig.~\ref{fig:CorrDec}(b) (data for $N=18$ and for
the less symmetric cluster with $N=36$ sites are discarded when fitting) yield the estimate $C^{\infty}_{\rm Plaq./Col.} \sim 0.16$, close to the value
$9/50=0.18$ expected for the pure {\em plaquette} VBC state (see the Appendix \ref{sec:PureStates}).

\begin{figure}[!htb]
\begin{center}
\includegraphics[width=\linewidth]{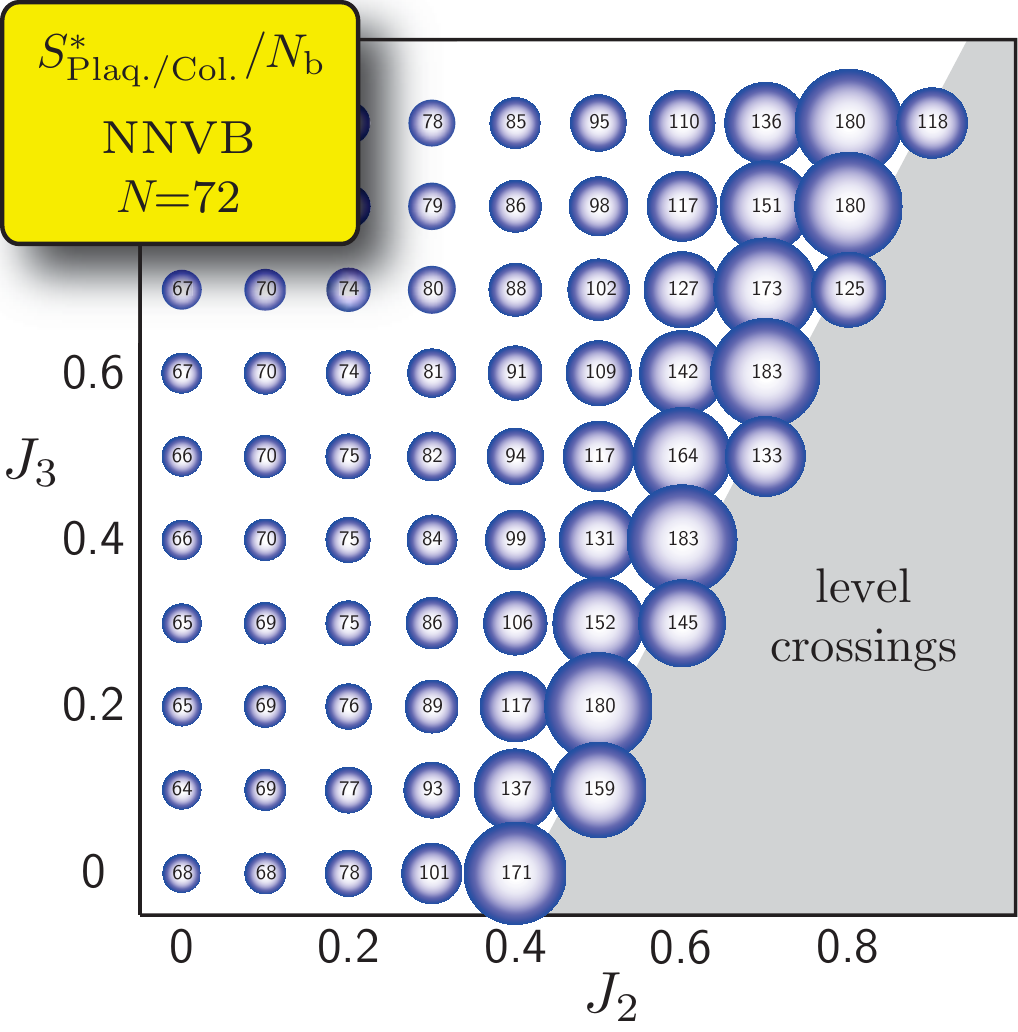}
\caption{(Color online) $S_{\rm Plaq./Col.}^{\ast}/N_{\rm b}$ for $N=72$, as obtained by solving the GEP Eq.~(\ref{eq:GEP}) in the NNVB subspace, as a function of $(J_2,J_3)$. The radius of the circles is proportional to $S_{\rm Plaq./Col.}^{\ast}/N_{\rm b}$. Numbers correspond to $10^3 S_{\rm Plaq./Col.}^{\ast}/N_{\rm b}$.
\label{fig:NNVB_dimer_structure_factors}}
\end{center}
\end{figure}

Finally, we analyze the strength of VBC order throughout the parameter space and in Fig.~\ref{fig:NNVB_dimer_structure_factors} we plot
$S_{\rm Plaq./Col.}^{\ast}/N_{\rm b}$ for the $N=72$ site cluster as a function of $J_2,J_3 \in [0,1]$. Unlike what happens for smaller clusters, for which
correlations mismatching the sign structure of the {\em columnar/plaquette} VBC patterns are found in part of the parameter space, four-point correlations for the $N=72$ cluster are
always fully consistent with {\em plaquette} VBC order up to the point where, for given $J_3$  and increasing values of $J_2$, one enters a regime
(highlighted in Fig.~\ref{fig:NNVB_dimer_structure_factors}) signaled by the occurrence of successive ground state level crossings (see Fig.~\ref{fig:SpecN54_VBxQDM} left panel for $J_3=0.3$), likely associated with the
breakdown of a description solely in terms of NNVB states (see related discussion in App.~\ref{sec:comp_NNVB_QDM}). Maximal values of $S_{\rm Plaq./Col.}^{\ast}/N_{\rm b}$ 
are observed just before the first such level crossing occurs. For comparison we refer to  the same quantity obtained by ED in the $S^z$ using the $N=24$ cluster in the left panel of 
Fig.~\ref{fig:NNVB_dimer_structure_factors_ed}. For larger $J_3$ values the agreement is quite remarkable, however as $J_3$ is reduced to zero, the VBC correlations are significantly
reduced in the ED approach compared to the NNVB results.

On the other hand, the region signalled on Fig.~\ref{fig:NNVB_dimer_structure_factors_ed} (right panel) by a strong {\em staggered} signal, does not appear to occur in a parameter range where the NNVB method can be safely used due to a rather modest variational energy (see Fig.~\ref{fig:E0_Error}) and the successive ground state level crossings (see Fig.~\ref{fig:NNVB_dimer_structure_factors} and Fig.~\ref{fig:SpecN54_VBxQDM} left panel for $J_3=0.3$). It may seem surprising that a {\em staggered} VBC state fails to be naturally captured by the NNVB approach. However two distinct arguments can explain this paradoxical situation. (i) As mentionned in section~\ref{subsec:ed_dimercorrs}, no definitive evidence supports the fact that the ground state is a singlet (non magnetic) state at the thermodynamic limit. In this respect a divergence of $S_{\rm Stag.}$ is only a signal of the spatial symmetry breaking associated to it, but does not preclude the possibility of a (magnetic) nematic phase which would obviously be out of reach of the NNVB method. (ii) If the ground state is a dressed {\em staggered} (singlet) state, a  good variational energy in the NNVB scheme may be hard to reach due to the structure of the {\em staggered} NNVB configuration (depicted in Fig.~\ref{fig:StateDefinitions} (c)) : contrary to the case of the {\em columnar} or {\em plaquette} states (see Fig.~\ref{fig:StateDefinitions} (a) and (b)), resonating NNVB configurations needed to dress the pure VBC state and lower its variational energy involve long resonating loops in the corresponding overlap diagram (see App.~\ref{sec:PureStates}) hence producing exponentially small overlaps and corrections to the bare VBC energy. In summary, this effective locking of the staggered NNVB configuration could make it difficult to emerge in the NNVB approach. A route to cure this issue and allow a more efficient relaxation to other VB configurations may be to include longer range dimer configurations assuming that the GEP remains numerically tractable.  

\subsection{Exact Diagonalization of an Effective Quantum Dimer Model}
\label{sec:EDQDM}

The NNVB approach used in the previous section requires an extensive
numerical treatment of the non-orthogonality of the VB states. It is
therefore numerically demanding and precludes the use of efficient
iterative algorithms, such as the Lanczos~\cite{Lanczos1950} algorithm. In this
respect, it would be desirable to base the study on a reliable (orthogonal) quantum
dimer model (QDM) in order to significantly increase the accessible system sizes.

Recently, a generic scheme for the derivation of QDMs from underlying Heisenberg Hamiltonians has been 
proposed in the context of two-dimensional frustrated antiferromagnets~\cite{Ralko2009,Schwandt2010}. This method 
aims to transform the generalized eigenvalue problem of the Heisenberg model in the short range valence 
bond basis (which was discussed in the preceding subsection~\ref{subsubsec:NNVB}) to an effective orthogonal
eigenvalue problem. In practice the transformation is conveniently carried out by a truncated diagrammatic 
expansion, containing only the most relevant terms. This derivation is presented in App.~\ref{sec:derivationeffQDM}, 
and provides a rather simple quantum dimer model with single- and double-hexagon resonance and potential terms, that reads
\begin{align}
{\cal H}_{\text{eff}}= & -t_6\hon{2_1}-t_{10}\hon{4_2} +\frac{t_{10}}{8}\hon{4_5} \nonumber\\
& +\frac{t_6}{4}\hon{2_2}+v_{10}\hon{4_6}-\frac{v_{10}}{2}\hon{4_7}~.\nonumber\\
~\label{eq:EffQDM}
\end{align}
%
Interestingly, the parameters of the effective QDM (given in App.~\ref{sec:derivationeffQDM}) only depend on the ratio $J_2/(J_1+J_3)$, so that the physics is 
invariant along simple isolines in the $J_2-J_3$ plane within the QDM description. In retrospect - 
comparing to Fig.~\ref{fig:NNVB_dimer_structure_factors} --- this invariance is also exhibited approximately by the 
NNVB approach, and within the boundaries of the VBC phase also to some extent in the $S^z$-ED 
approach~(cf.~left panel of Fig.~\ref{fig:NNVB_dimer_structure_factors_ed}).

The leading contribution of the QDM, with terms on a single hexagon,
is a simple Hamiltonian of the Rokshar-Kivelson form~\cite{rokhsar:88}, 
with off-diagonal and diagonal terms with amplitudes $t$ and $V$ respectively, 
that has been studied 
in great detail in Ref.~\onlinecite{Moessner2001a}. 
It turns out that the ratio $V/t=1/4$ is fixed (independent of $J_2$ and $J_3$), for which case 
it was shown that the ground state is in the {\em plaquette} phase. Since our effective
Hamiltonian reduces to this form when $J_2/(J_1+J_3)=3/8$ (i.e. the two-hexagon terms vanish), 
we expect that {\em plaquette} physics will occur in this region. However, it is not yet
clear what will be the extent of this phase, since when we move away from this line, the 2-hexagon
terms appearing in the effective QDM might alter this behavior.

\begin{figure}[!htb]
\begin{center}
\includegraphics*[width=\linewidth]{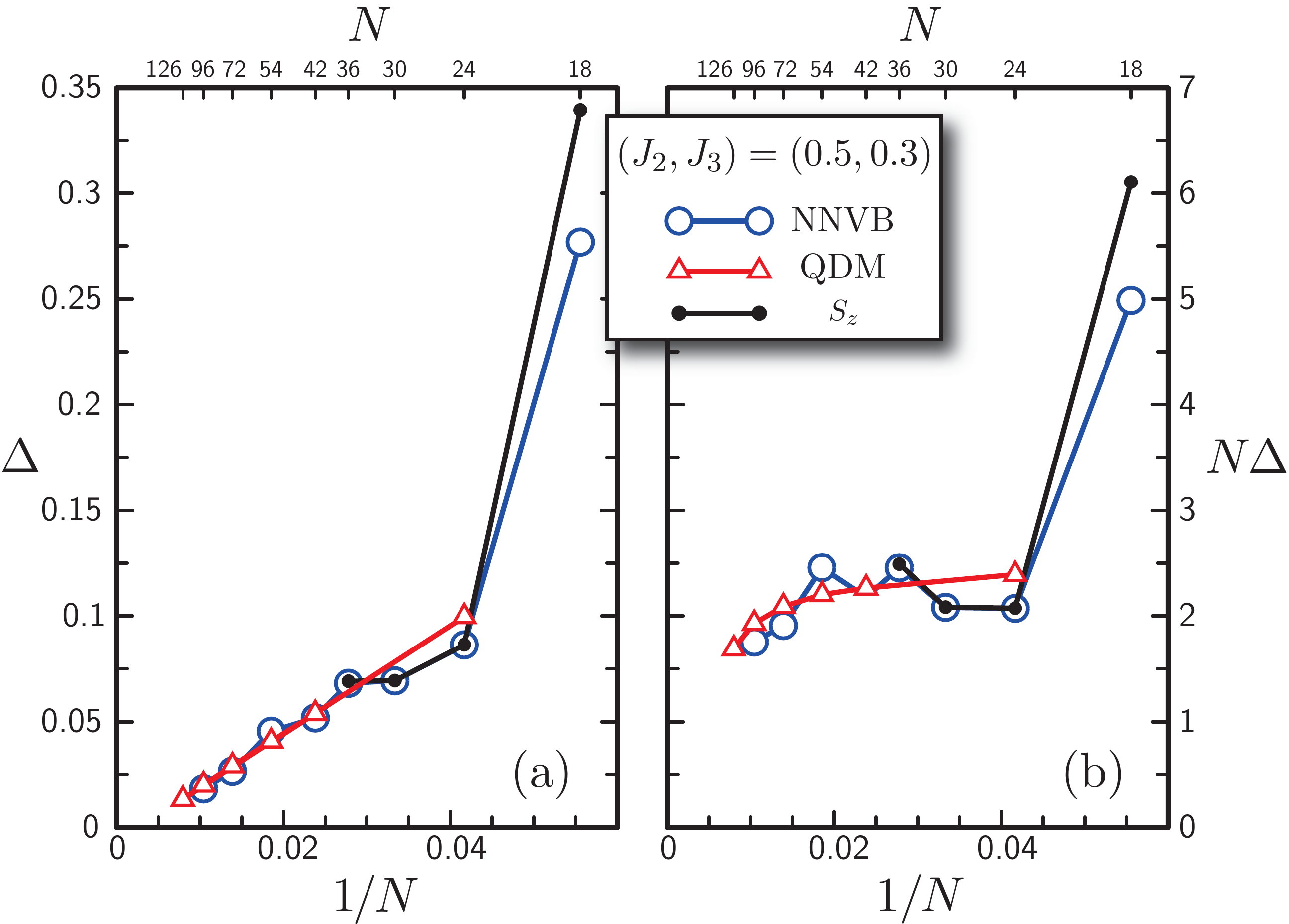}
\caption{(Color online) Finite-size scaling of the singlet gap between the ground-state and the lowest excitation with momentum $K$ for $J_2=0.5$ and $J_3=0.3$, obtained
by ED in the $S^z$ basis, the NNVB and the effective QDM approach. Note that the singlet gap matches quite well between the different techniques, despite the somewhat
poor variational energy of the NNVB and QDM approach. 
\label{fig:Gap}}
\end{center}
\end{figure}

\subsubsection{Comparison with NNVB}

In deriving the QDM in App.~\ref{sec:derivationeffQDM} we made several assumptions and simplifications. First of all we neglected subdominant terms in order to keep the model simple. This may cause some approximation errors that can be detected by comparing the QDM with NNVB approach. Secondly, the QDM is derived for an infinite lattice~\cite{Schwandt2010}, which substantially improves the finite size scaling behavior. As shown in App.~\ref{sec:comp_NNVB_QDM} the ground state energy converges to the same value as the NNVB approach, however for the QDM the convergence is much faster, i.e. it has smaller finite size corrections. This validates the quantum dimer model, and also explains the finite size differences between the QDM and the NNVB approach.

A careful comparison between both approaches is presented in Appendix~\ref{sec:comp_NNVB_QDM}. In order to illustrate on a specific example how all these methods agree, we have computed the finite-size gap to the first singlet excitation with momentum $K$ for $(J_2,J_3)=(0.5,0.3)$. This set of parameters was chosen based on ED correlations computed for an $N=24$ cluster (with a grid spacing of 0.1), since it provides a strong {\em plaquette} structure factor $S_{\rm VBC}$ (see Eq.~\ref{eq:structure} for its definition). As has been discussed already, both {\em plaquette} and {\em columnar} VBC have the same discrete symmetry breaking, corresponding to a 3-fold degeneracy in the thermodynamic limit. For increasingly larger cluster sizes, the lowest singlet excitation at the two equivalent $K$ points should collapse onto the ground-state. Moreover, because of the finite correlation length in the VBC, this singlet gap must ultimately vanish exponentially with increasing system size. 

In Fig.~\ref{fig:Gap}, we plot the scaling of this singlet gap $\Delta$ 
computed with all our numerical techniques. Unbiased ED data already shows a clear indication of a vanishing gap in the thermodynamic limit, but the scaling seems rather $\sim 1/N$ since we cannot reach large
enough cluster sizes.  Using NNVB data, we can extend our computations to larger clusters (up to $N=96$), and we observe 
an excellent agreement when comparing to ED data. This is not obvious since for instance the variational NNVB energy is not that accurate (see Fig.~\ref{fig:E0_Error} and App.~\ref{sec:comp_NNVB_QDM}), but computing energy differences can give accurate results when there is a systematic deviation in all energies. Looking at the scaling of NNVB data, we see a behavior that could be compatible with a scaling faster than $1/N$, but there are still some irregularities in the finite-size effects due to different cluster shapes. Our last set of data is obtained from the QDM model, that was simulated up to a $N=126$ cluster: we note a semi-quantitative agreement with other techniques for the gap numerical data. Moreover, as explained above, the QDM has much weaker finite-size effects, which is clearly observed in the plot where finite-size scaling is much smoother. The possibility to access large clusters with small finite-size effects allows us to show convincingly that the gap collapses fast enough and that the system has long-range VBC order in the thermodynamic limit.

However, it remains to be determined which of the two potential VBC candidates ({\em columnar} versus {\em plaquette}) is realized. 

\begin{figure}[!ht]
\begin{center}
\includegraphics[width=\linewidth]{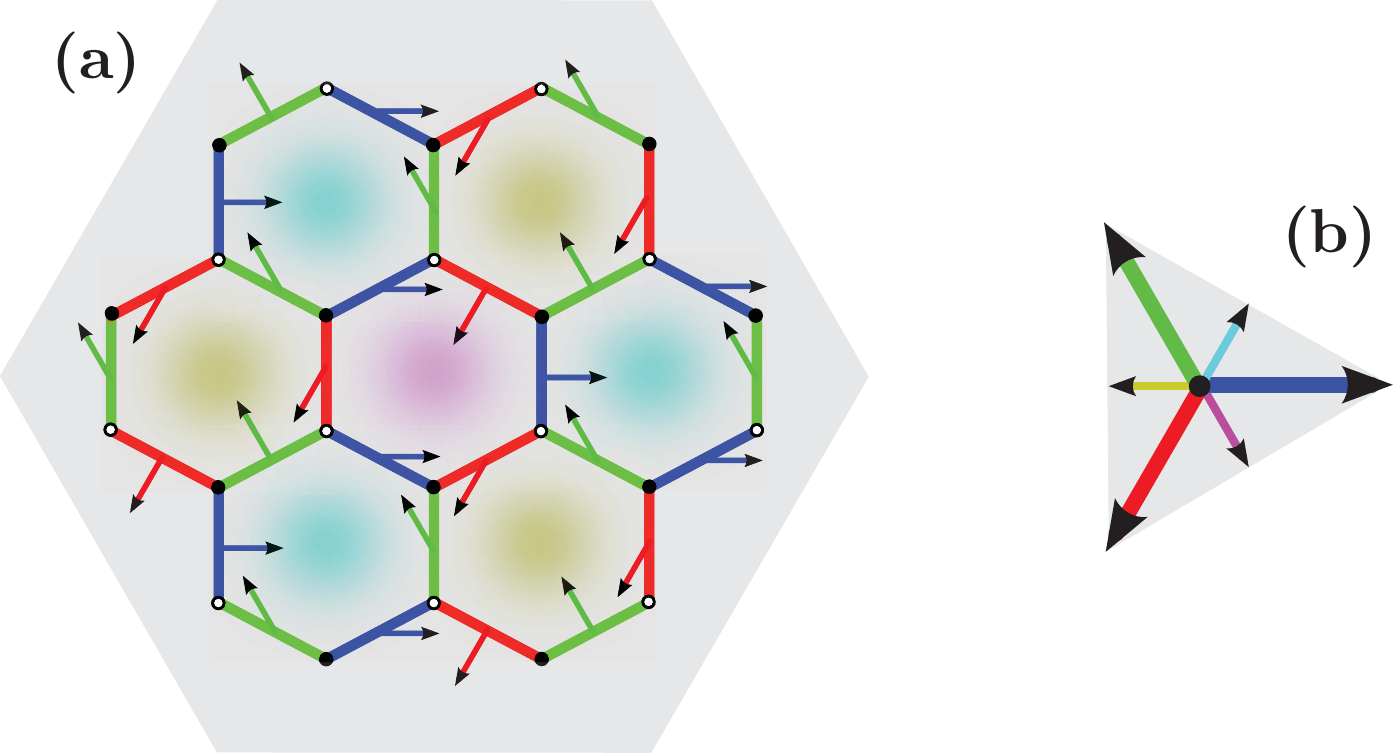}
\caption{(Color online) (a) all dimers belonging to the same of three possible {\em columnar} states have the same dimer vector associated. A resonating {\em plaquette} contains two different dimer vectors, that contribute equally to the resulting histogram. (b) the phase space built by the dimer vectors forms an equilateral triangle, where the corners represent the {\em columnar} states, while the {\em plaquette} states are signaled by a binomial distribution on the edges.
\label{fig:dvh}}
\end{center}
\end{figure}

\begin{figure*}[!ht]
\begin{center}
\includegraphics[width=0.8\linewidth]{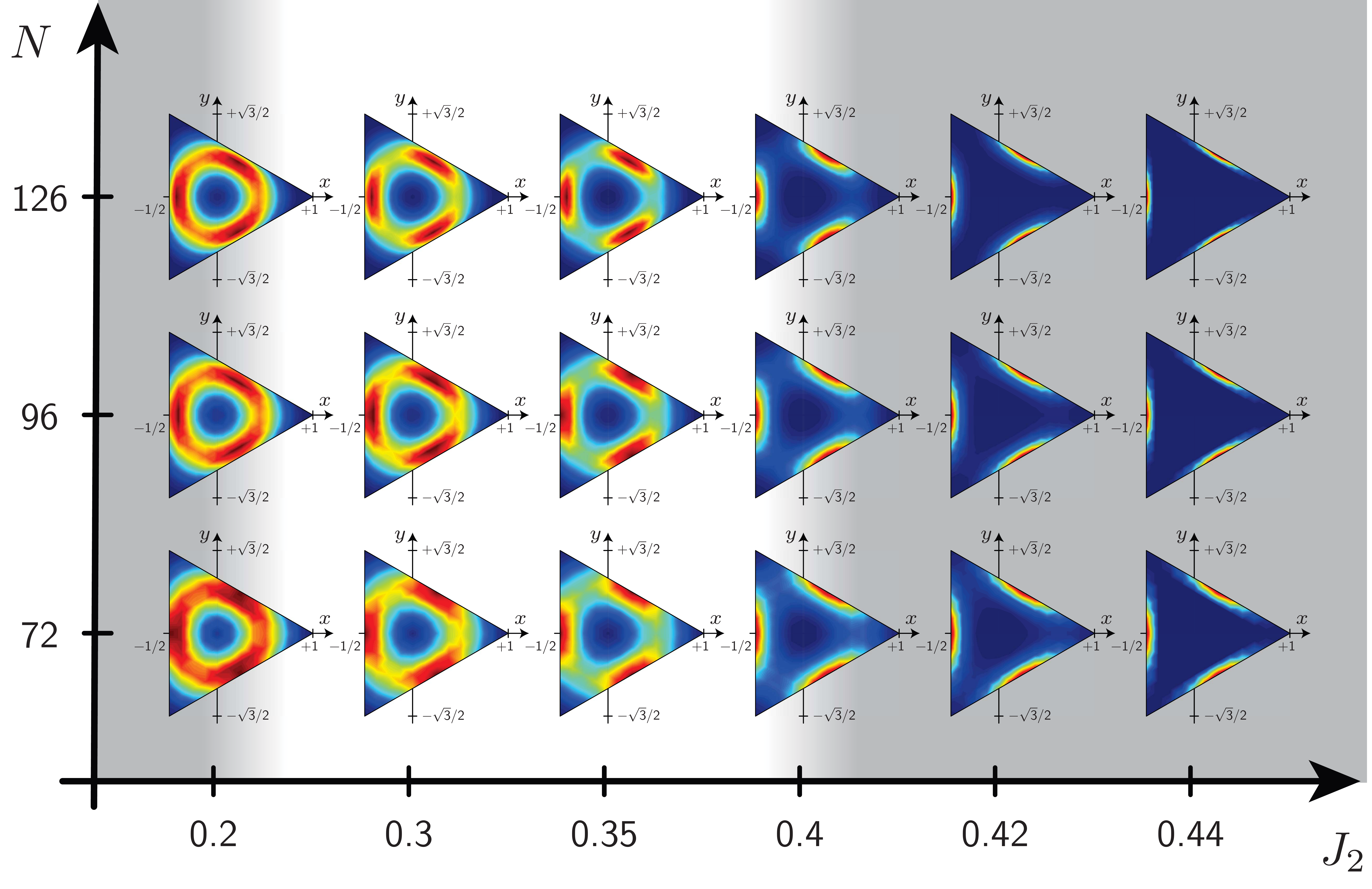}
\caption{(Color online) Normalized dimer histograms $P(N_x,N_y)$, as defined in Eq.\eqref{eq:dimer_histo}, obtained within the effective QDM for different system sizes at $J_3=0$. We
observe pronounced {\em plaquette} VBC signals at the larger $J_2$ values shown, with a tendency towards a reduced radius and a more $U(1)$-symmetric behavior as the N\'eel phase at 
smaller $J_2$ is approached.
\label{fig:DH}}
\end{center}
\end{figure*}

\subsubsection{Dimer Vector Histograms}

Studying (orthogonal) quantum dimer models offers two advantages: first of all one can study larger systems. Here for example we were able to study honeycomb samples with up to $N=126$ sites, the
second, even more interesting point is that one has access to new observables which are hard to define and implement in either the $S^z$ or the NNVB basis. In order to detect the underlying phase of a Hamiltonian, one usually measures correlations in the ground state, as done in Sec.~\ref{sec:fourpoint}. However, the QDM allows for the computation of a
related useful observable~\cite{sandvik:07,Kawashima:07,Misguich2008b,Lou2009,Beach:09}. The idea is to associate a two-dimensional vector to every dimer and collect a histogram of the vector occupations.
Writing the ground state $\ket{\psi_0}=\sum_i a_i \ket{\varphi_i}$ as superposition of orthogonal dimer configurations $\ket{\varphi_i}$, one defines the appropriate histogram as
\begin{equation}
P(N_x,N_y)=\sum_{i\in{\cal C}_{N_x,N_y}}  \left|a_i\right|^2,
\label{eq:dimer_histo}
\end{equation}
where ${\cal C}_{N_x,N_y}$ is indexing all dimer states $\ket{\varphi_i}$ that have a total dimer vector
\begin{equation}
(N_x,N_y)=\sum_{[i,j]\in\ket{\varphi}} {\bf v}_{[i,j]}.
\end{equation}

The left panel of Fig.~\ref{fig:dvh} illustrates a particular choice of dimer vectors, that assigns three different vectors to the three {\em columnar} (Read-Sachdev) states. The phase space of the resulting histogram 
forms a triangle illustrated in the right panel,  where the corners of the triangle represent the {\em columnar} states, while the {\em plaquette} states are signaled by a binomial distribution on the edges of the triangle. A
{\em staggered} VBC state on the other hand would contribute to the center of the histogram.

In Fig.~\ref{fig:DH} we display dimer vector histograms obtained within the QDM approach for the three largest samples with $N=72,96$ and $126$ sites, and for several $J_2$ values at
$J_3=0$. The parameter region in which the QDM approach is expected to be appropriate for the original spin model is highlighted with a white background, whereas the remaining parameter region
is shaded in grey. Let us start the discussion at $J_2=0.4$, which is close to the point $J_2=3/8=0.375$, where the QDM reduces to the Rokhsar-Kivelson model at $V/t=1/4$, expected to
display a {\em plaquette} VBC ground state~\cite{Moessner2001a}. Indeed the histogram displays pronounced peaks in the middle of the edges of the triangle, as expected for a {\em plaquette} VBC phase.
These results thus corroborate the complementary findings based on four-point correlation functions using ED in the $S^z$ basis and the NNVB approach.
Pushing the QDM somewhat further into the ``unphysical'' region of larger $J_2$, the {\em plaquette} signal is even more pronounced. On the other hand by lowering $J_2$ towards $J_2 \sim 0.2$, the histogram
becomes more rounded and fuzzy, reminiscent of the emergent $U(1)$ symmetry at deconfined quantum critical points at N\'eel to VBC transitions~\cite{Senthil04,Senthil04b,sandvik:07,Lou2009,Beach:09}. 
Upon lowering the $J_2$ parameter, the radius is also somewhat reduced and the dimer correlation length increases, however our effective QDM using nearest-neighbor
valence bonds only is not able to reproduce the vanishing of the VBC order parameter (i.e. the radius of the distribution) as the N\'eel phase is approached. In spite of this
limitation, the approximate $U(1)$-like symmetry exhibited by the histograms upon approaching the N\'eel phase may well be a physical feature of the N\'eel to VBC transition on the 
honeycomb lattice.

\section{Conclusion}
\label{sec:conclusion}

In the present work we have analyzed the phase diagram of the frustrated
$J_1-J_2-J_3$  spin-1/2 Heisenberg model on the honeycomb lattice by
using a combination of different ED approaches and a SCMFT treatment.
We have localized the boundaries of several magnetically ordered phases
in the the region $J_2,J_3\in [0,1]$, and found a sizable magnetically disordered region in between. We characterize
a large part of this magnetically disordered region as a {\em plaquette} valence bond crystal phase. 
Interestingly we find that a particular parameter-free Gutzwiller projected tight-binding wave function
has remarkably accurate energies compared to finite-size extrapolated ED energies along the transition line
from the well-known N\'eel phase to the {\em plaquette} VBC, a fact that points to possibly interesting critical behavior - such as deconfined criticality -
across the transition. In contrast a direct N\'eel to {\em staggered} VBC transition has recently been shown to be strongly first order~\cite{Banerjee2010}.

Compared to previous work on the $J_1-J_2-J_3$ phase diagram we localize precisely the magnetic phases (phases I and II in the phase diagram shown in 
Fig.~\ref{fig:phase_diagram}) which have been discussed to be present at the semiclassical level~\cite{Rastelli19791,Mattsson1994,Fouet2001,Mulder2010,Cabra2010,Mosadeq2010},
and we discuss the possibility of a reentrant collinear magnetic phase IV in a region at larger $J_2$, which would nevertheless
be compatible with the {\em staggered} dimer correlations found earlier in the relevant region~\cite{Fouet2001,Mulder2010,Mosadeq2010}. The
possibility of a {\em plaquette} phase has been discussed recently in Ref.~\onlinecite{Mosadeq2010} restricted to the $J_3=0$ line, while an earlier work~\cite{Fouet2001}
reported that the dimer correlations might not be sufficiently strong for a plaquette VBC, and put forward the idea of an RVB liquid. 

Here we established that a plaquette phase does indeed occur for larger $J_3$ values when leaving the N\'eel phase (I) by computing dimer-dimer 
correlations via exact diagonalizations in the NNVB subspace and the analysis of dimer histograms within an effective QDM, and we showed that the 
phase has a sizable extent in the $J_3$ direction, including the magnetically disordered region found recently on the $J_2=J_3$ line~\cite{Cabra2010} (and thereby clarifying its nature). 
The precise fate of the plaquette VBC upon approaching the $J_3=0$ line is however still an open question.

The situation regarding the staggered VBC versus magnetic order in phases III and IV in Fig.~\ref{fig:phase_diagram} is  not clear yet, and the possibility of incommensurate
behavior of spin correlations or magnetic order renders an ED analysis quite challenging. It is likely that this question can be more meaningfully
addressed using coupled-cluster or spin f-RG techniques, as done recently in the context of incommensurate spin correlations on the frustrated square lattice~\cite{Reuther2011}.

At the technical level it is notable that we have found an interesting example where we could explicitly
show that it is possible to derive an effective quantum dimer model which accurately describes
the magnetically disordered {\em plaquette} VBC region. Such a connection was conjectured to be
present already some time ago~\cite{Moessner2001a}. However, no precise connection between 
a QDM and original spin models could be made at that time. It is still an open question to understand
why the NNVB and the QDM approach are currently unable to detect and describe the {\em staggered}
VBC (lattice nematic) discussed previously. It might be that both methods are biased towards dealing
with valence bond configurations which retain some flipability on short loops, while the {\em staggered} VBC
configurations do not contain short flipable loops at all. On the other hand it could also be that the
lattice nematic state actually is a magnetically ordered state (at least in some part of parameter space)
which breaks the same lattice symmetries as the the {\em staggered} VBC, giving rise to qualitatively similar 
dimer-dimer correlations.

Returning to one of the initial motivations --- the understanding of the magnetism of the half filled Hubbard model upon lowering of $U/t$, and
the possible explanation of the spin liquid behavior found in Ref.~\onlinecite{Meng2010}  ---
 the questions are: i) what is the effect of the sub-leading next nearest neighbor $J_2$ correction to the nearest neighbor $J_1$ Heisenberg
interaction in terms of new phases arising;  and ii) are the required values of $J_2/J_1$ for new physics beyond the N\'eel phase reachable 
by downfolding the Hubbard model to a spin model at intermediate $U/t$, or does one have to consider more correction terms?

Regarding i): the scenario developed in the present paper is that $J_2$ destabilizes N\'eel order somewhere between $J_2/J_1\sim0.17-0.22$, and then
a {\em plaquette} VBC phase (or a disordered version thereof) sets in, up to a value of $J_2/J_1\sim 0.35-0.4$. For even larger values of $J_2/J_1$ a lattice nematic ({\em staggered} VBC) [c.f. Refs.~\onlinecite{Fouet2001,Mulder2010,Mosadeq2010} 
and right panel of Fig.~\ref{fig:NNVB_dimer_structure_factors_ed}] as well as magnetically ordered spiral phases can arise. 
Based on the success of the Gutzwiller projected Dirac sea wave function to 
quantitatively describe the energies at the N\'eel to {\em plaquette} VBC transition, as well as the qualitative agreement with respect to spin-spin and dimer-dimer
correlation functions, we suggest that a deconfined critical point scenario might describe this particular N\'eel to VBC transition on the honeycomb lattice.
Our current ED tools are admittedly not perfectly suitable to resolve more complex scenarios, such as an SU(2) algebraic spin liquid region~\cite{Hermele2007} with a small
but finite extent. The same is true for a small $\mathbb{Z}_2$ spin liquid region~\cite{Wang2010,LuRan2010,LuRan2010_b,Vaezi2010,Xu2011} appearing 
between the N\'eel and {\em plaquette} phase. We also note that a recent instanton analysis of one kind of $\mathbb{Z}_2$ spin liquid revealed an instability to
a VBC phase~\cite{Vaezi2011}, in agreement with the {\em plaquette} VBC phase we find. 
Our analysis is however at variance with the phase diagram put forward in the VMC study of Ref.~\onlinecite{Clark2010}. In that work the succession of phases is 
N\'eel, a rather large $\mathbb{Z}_2$ spin liquid region, followed by a rotational symmetry breaking state.

Regarding the question ii) a recent estimate on the ratio of $J_2/J_1$ in the spin liquid region of the honeycomb Hubbard model was put forward based on continuous unitary transformations
in Ref.~\onlinecite{Yang2010b}, and a value of about $J_2/J_1\sim 0.06$ was quoted. While this value seems to be almost sufficient to enter a new phase in the $J_1-J_2$ model according to 
the VMC analysis of Ref.~\onlinecite{Clark2010}, which reported a critical value of $J_2/J_1 = 0.08$, our ED based value for the critical ratio is at around three times as large ($0.17-0.22$). We currently 
believe that the small $J_2/J_1$ value in Ref.~\onlinecite{Clark2010} is due to a comparatively poor variational energy of the N\'eel state, when compared to our finite size extrapolated ED energies, 
therefore shifting the VMC transition to a too small $J_2/J_1$  value. So we believe based on our results that a simple $J_1-J_2$ spin model alone does not allow a {\em quantitative} description of the spin 
liquid phase discovered recently in the Hubbard model~\cite{Meng2010}. More work is needed to understand whether the phase adjacent to N\'eel phase at $J_3=0$ is a plaquette VBC with a small order parameter
or a genuine spin liquid phase, in which case the $J_1-J_2$ model at small $J_2$ would at least qualitatively explain the physics of the Hubbard model on the insulating side of the Mott transition. Future efforts will also have to explore the effects of higher order corrections and thereby unravel whether a quantitative spin-only description of the spin liquid phase in the Hubbard model on the honeycomb lattice is possible.

{\em Note added}: After submission of this work we became aware of Ref.~\onlinecite{Reuther2011b}, where a spin f-RG study of the
same model is presented. In that paper a considerably large magnetically disordered phase is also found.

\acknowledgements

The authors thank B. Clark, M.~Hermele, R.~Moessner, R.~Thomale and A.~Vishwanath for useful discussions and C. Weber
for correspondence on the variational Monte Carlo results. We also acknowledge interesting discussions with 
Z.Y.~Meng, T.C.~Lang, S.~Wessel, F.F.~Assaad and A.~Muramatsu on the results of Ref.~\onlinecite{Meng2010}.
AML is grateful to H.-Y.~Yang and K.P.~Schmidt for collaboration on related topics. SC and AML acknowledge the KITP for hospitality 
in the final stage of the project and support through NSF grant No PHY 05-51164. DS thanks the MPI PKS for support through its visitors 
program. This work was performed using HPC resources from MPG RZ Garching, GENCI-IDRIS (Grant 2009-100225) and CALMIP. 
This work is supported by the French ANR program ANR-08-JCJC-0056-01.


\begin{figure}[!htb]
\begin{center}
\includegraphics[width=0.9\linewidth]{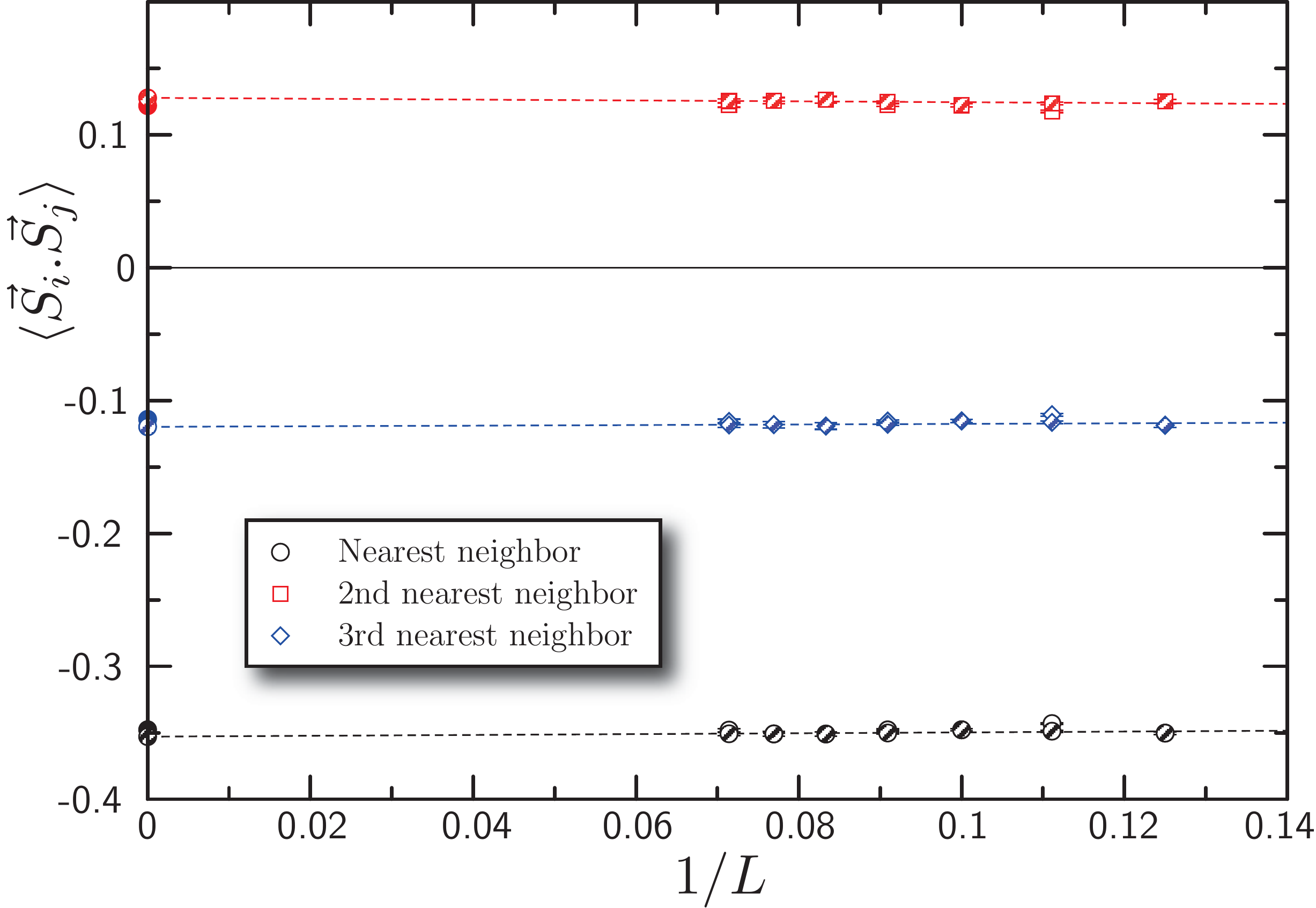}
\caption{(Color online) Gutzwiller projected Dirac sea: 
Finite size scaling behavior and extrapolation $L\rightarrow\infty$
of the spin correlations relevant the energy of the frustrated honeycomb 
Hamiltonian (\ref{eqn:HeisenbergHamiltonian}). See text for the explanation of the
two types of symbols shown.}
\label{fig:GPDS_local_energies} 
\end{center}
\end{figure}

\appendix

\section{Properties of the Gutzwiller-Projected Wave Function}
\label{sec:prop_gutzwiller}

Wave function approaches to strongly correlated systems can be valuable
approximations because a good variational wave function can provide significant 
physical insight due to its relative simplicity. In the present study we focus on a 
(completely) Gutzwiller projected half-filled nearest neighbor hopping tight binding 
model on the honeycomb lattice. As this corresponds to a filled Dirac sea we term
the wave function Gutzwiller projected (GP) Dirac Sea~\footnote{This wave function 
has also been studied independently in~Ref.~\onlinecite{Clark2010}, where it served 
as the prototype wave function for an algebraic spin liquid (ASL).}.

Here we use a standard Monte Carlo procedure to evaluate correlation functions of the
parameter free wave function according to the update scheme proposed by Ceperley,
Chester and Kalos~\cite{Ceperley1977}. While doing so we noticed the occurrence of 
slowly equilibrating starting configurations which had a significant effect in determining 
some of the correlation functions and their error bars. It is presently not clear to us whether
this is due to a inefficient Monte Carlo sampling or due to a fat tailed distribution for
some observables, as for example discussed in Ref.~\onlinecite{Trail2008} for continuum 
systems.

First we determine the nearest neighbor, next-nearest and third-nearest neighbor spin-spin
correlation function, as they allow us to determine the variational energy of this wave function
for the $J_1{-}J_2{-}J_3$ Heisenberg Hamiltonian [Eq.~(\ref{eqn:HeisenbergHamiltonian})] studied
in this paper. The finite size expectation values for lattices with $N=2\times L^2$ with $L=8,\ldots,14$ are
displayed in Fig.~\ref{fig:GPDS_local_energies}. Two different data sets are shown, first the bare
estimates with error bars including all independent Markov chains (open symbols), and a second set where 
the anomalous samples were removed in calculating the mean and the error bars (hatched symbols). The 
later procedure yields estimates which show a markedly smoother finite size behavior and
are used to linearly extrapolate the estimates to $L\rightarrow \infty\ $~\footnote{The system sizes $L$ which are multiples of three
have been discarded, because the open shell conditions complicate the analysis.}.  The energy per site is then found to be approximately:
\begin{eqnarray}
E(J_1,J_2,J_3)/N&\approx& -0.353 \times 3/2 \times J_1\nonumber\\
&&+ 0.128 \times 3 \times J_2\nonumber\\
&&-0.120 \times 3/2 \times J_3\ .
\label{eqn:dirac_energy}
\end{eqnarray}
As already shown in Fig.~\ref{fig:gs_energy} (for $J_3=0$ in the left panel and $J_3=0.3$ on the right panel), the energy per site
of this wave function is very close to the extrapolated energy per site of the frustrated Heisenberg 
model close to the supposed N\'eel to {\em plaquette} transition. This success is quite striking for a parameter free wave function.

\begin{figure}[!htb]
\begin{center}
\includegraphics[width=0.9\linewidth]{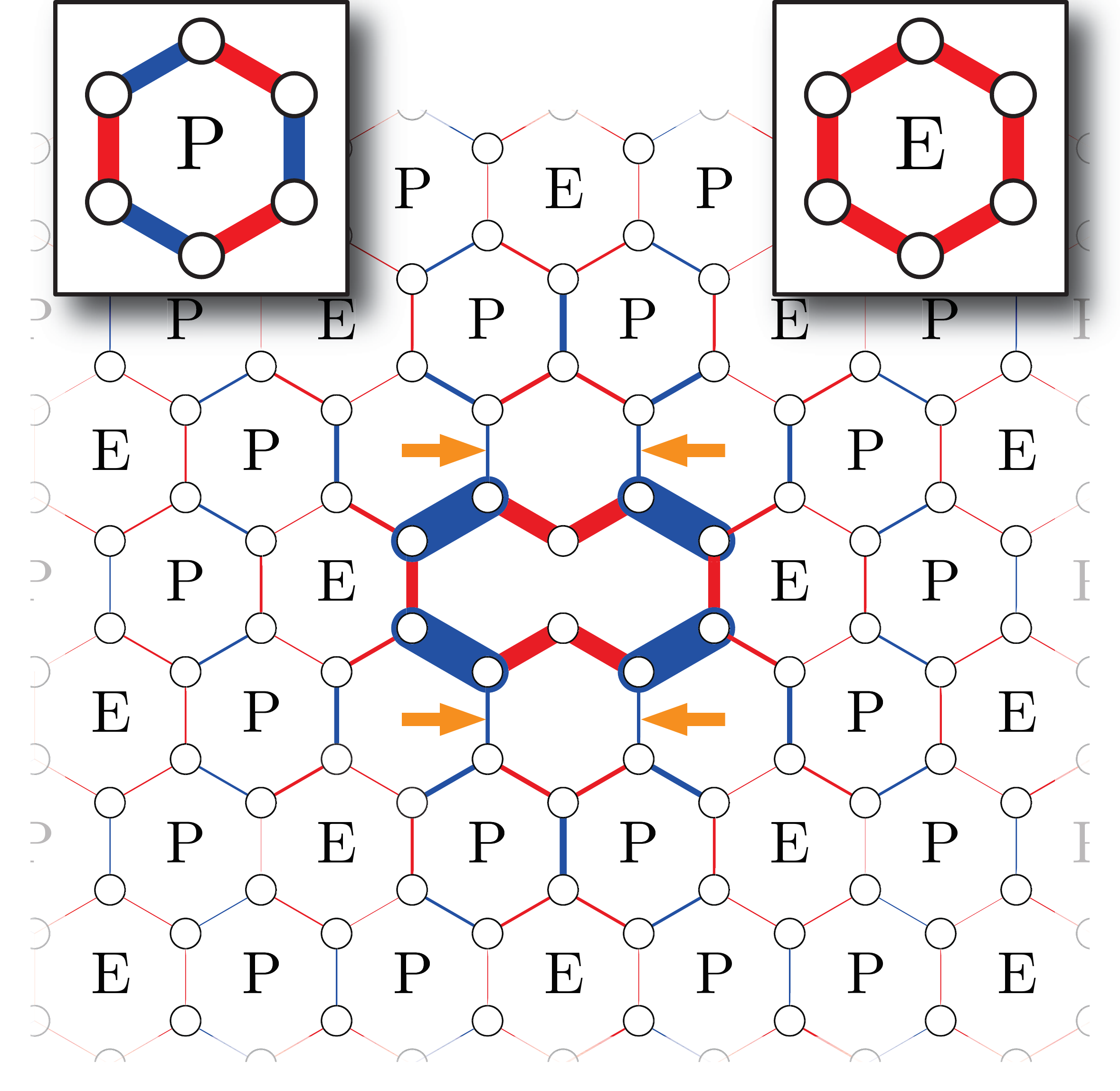}
\end{center}
\caption{(Color online) Dimer correlations [Eq.~\ref{eqn:sz_dimer_corrs}] evaluated in the
Gutzwiller projected Dirac sea for a $L=11$ sample. The black bond denotes the reference
bond, while the blue (red) bonds denote positive (negative) correlations. The width of the bonds
are proportional to the value of the correlation function. The (E) letter indicates hexagons with
all negative correlations, while (P) indicates plaquettes with a staggered signal. The
four bonds indicated by arrows are the only ones that differ in sign from the expectations for
a Read-Sachdev or {\em plaquette} VBC.}
\label{fig:GPDS_dimer_correlations}
\end{figure}
This surprising observation raises the question whether the wave function exhibits appropriate correlations 
to describe such a transition beyond the competitive ground state energy. We have therefore determined the spin correlation 
functions at larger distances and found that correlations are perfectly staggered according to the N\'eel pattern
and decay algebraically as $1/r^{\sigma_s}$ with a decay exponent $\sigma_s \approx {1.7(2)}$. Next we
have measured dimer-dimer correlations functions of nearest neighbor bonds
\begin{equation}
C^{zz}_{ijkl}=\langle(S^z_i S^z_j)(S^z_k S^z_l)\rangle - \langle S^z_i S^z_j\rangle \langle S^z_k S^z_l\rangle \ ,
\label{eqn:sz_dimer_corrs}
\end{equation}
and display the correlation pattern for an $L=11$ system in Fig.~\ref{fig:GPDS_dimer_correlations}.
As already pointed out in Ref.~\onlinecite{Clark2010}, the short range structure of the dimer correlations in this
wave function is surprisingly analog to the one found in the spin liquid phase of the honeycomb Hubbard
model~\cite{Meng2010}. What has however not been noticed previously is that beyond the four "inverted" 
bonds (highlighted by arrows in Fig.~\ref{fig:GPDS_dimer_correlations}) all the other dimer correlations 
explored here match the signs expected for {\em columnar} (Read-Sachdev) or {\em plaquette} VBC states as derived in 
App.~\ref{sec:PureStates}. The dimer correlations also seem to decay as a power law, but we have not been able to 
determine the corresponding decay exponent accurately enough. 

The correlations measured in the Gutzwiller projected Dirac sea qualify this wave function as a viable candidate
to describe a critical state separating a N\'eel ordered magnetic phase from a VBC of {\em columnar} (Read-Sachdev) or {\em plaquette}
type. The fact that this wave function simultaneously exhibits staggered N\'eel fluctuations, as well as {\em columnar/plaquette} VBC fluctuations,
is reminiscent of the SU(2) algebraic spin liquid state on the honeycomb lattice put forward by Hermele~\cite{Hermele2007}, which is however 
believed to describe an extended spin liquid region, instead of a single critical point~\cite{Hermele2005}. 
Further work is required to understand whether this wave function could possibly also represent a deconfined quantum 
critical point~\cite{Senthil04,Senthil04b} separating the two phases or whether there is indeed an extended algebraic spin liquid phase present between
the two ordered phases (N\'eel - {\em plaquette} VBC) discussed in this paper~\cite{Hermele2005}. 
Yet a different scenario has recently been advocated in Refs.~\onlinecite{Wang2010,LuRan2010,LuRan2010_b,Clark2010}, where a 
gapped $\mathbb{Z}_2$ spin liquid has been proposed as a phase neighboring the N\'eel ordered phase. Variationally the $\mathbb{Z}_2$ spin liquid
was found~\cite{Clark2010} to be a tiny fraction lower in energy than the GP Dirac sea studied above. Whether this is also true beyond the variational 
realm remains an open question.

\section{Derivation of an effective Quantum Dimer Model}
\label{sec:derivationeffQDM}

In this section we redefine the $J_1$--$J_2$--$J_3$--Heisenberg model on the honeycomb lattice (\ref{eqn:HeisenbergHamiltonian}) as
\begin{align}\label{eq:Heff}
{\cal H}_{\text{eff}}=\frac{4}{3}{\cal O}^{-1/2}{\cal H}{\cal O}^{-1/2}+\frac{N J_1}{2},
 \end{align}
where ${\cal H}$ is the matrix introduced in section \ref{subsubsec:NNVB}, ${\cal O}$ is the overlap matrix for the NNVB basis states and $N$ is the number of sites.

On the honeycomb lattice there is only one elementary process that resonates between the two possible valence bond coverings on a hexagon. As shown in Ref.~\onlinecite{Schwandt2010}, this naturally leads to a potential term, counting the number of flippable plaquettes. The exact amplitudes of both processes are shown to be given by $t_6=-(6J_2-3J_1-3J_3)\alpha^4/(1-\alpha^8)$ for the kinetic term and $v_6=t_6\alpha^4$ for the potential one. Here we choose the bipartite convention with $\alpha=1/\sqrt{2}$.

Interestingly, the amplitudes $t_6$ and $v_6$ depend only on one parameter, $J_2^\text{eff}=J_2/(J_1+J_3)$. Note, that this qualitatively agrees with the phase diagrams (Figs.~\ref{fig:phase_diagram} and \ref{fig:MF_phdiag}) suggested earlier in this paper. One can therefore simplify the Hamiltonian to
\begin{align}
{\cal H}_{\text{eff}}(J_1,J_2,J_3)=(J_1+J_3){\cal H}_{\text{eff}}\left(1,\frac{J_2}{J_1+J_3},0\right).
\end{align}

One can check easily, that this relation also holds for processes that connect dimer configurations defined on two hexagons. While those contributions cannot be obtained analytically, we find some iterative, numerical algorithm, that allows for calculating the amplitudes for all possible terms of this kind. This algorithm appears to converge rapidly and will be briefly described in the following.

While it is relatively easy to obtain the inverse of an operator within the present scheme~\cite{Schwandt2010}, calculating the square root is much less obvious. We therefore have to go beyond previous works in order to derive an expression for ${\cal O}^{-1/2}$. The idea is to explicitly work in a basis that is formed by all the diagrams that are considered. Hence, it is possible to write every sum of processes as a vector and every fusion as a linear map applied to this vector.
As an example, in the basis
\[\left\{\hon{2_1},\hon{2_2},\hon{4_2}\right\},\]
the fusion of two flips on a hexagon can be written as
\begin{equation*}
\left(\begin{array}{ccc}
0 & 1 & 0\\
1 & 0 & 0\\
1 & 0 & 0\\
\end{array}\right)\cdot
\left(\begin{array}{c}
1\\
0\\
0\\
\end{array}\right) = 
\left(\begin{array}{c}
0\\
1\\
1\\
\end{array}\right),
\end{equation*}
resulting in a contribution for a potential term on a hexagon and a kinetic term on two hexagons.\\

Generalizing and applying this procedure to a larger basis allows for an iterative solution of 
\[\frac{1}{2}\left\{{\cal O}^{-1/2},{\cal O}^{-1/2}\right\}={\cal O}^{-1},\]
to obtain ${\cal O}^{-1/2}$. Putting the result into \ref{eq:Heff}, we arrive at the Hamiltonian Eq.~(\ref{eq:EffQDM}),
with coefficients given by

\begin{equation*}
\begin{array}{ccp{0.5cm}cr}
\hline\hline
\multicolumn{2}{c}{\text{\bf 1 hexagon}} & & \multicolumn{2}{c}{\text{\bf 2 hexagons}} \\
\multicolumn{2}{c}{\left[\frac{4}{3}\left(2J^\text{eff}_2 - 1\right)\right]} & & \multicolumn{2}{c}{\left[\frac{4}{3}\left(8J^\text{eff}_2 - 3\right)\right]} \\[3pt]
\hline
t_6 & -0.6 &  & t_{10} & -0.049218(5)\\
 &  & & v_{10} & 0.001562(9) \\
\hline\hline
\end{array}
\end{equation*}

Note that $t_6$ changes sign at $J_2^\text{eff}=1/2$, while $t_{10}$ and $v_{10}$ change sign at $J_2^\text{eff}=3/8$. The ratio of $v_{10}/t_{10}$ does not depend on $J_2^\text{eff}$, although its analytical value is not known at the present stage. We note in passing that the model with only the most relevant $t_{10}$ term has been studied in the context of supersolids of hardcore bosons on the triangular lattice~\cite{Sen2008,Wang2009}, whereas the model at $t_{10}=v_{10}=0$ corresponds to a particular point of the Rokhsar-Kivelson model studied in Ref.~\onlinecite{Moessner2001a}.

The QDM combines the advantages from both exact diagonalizations in the $S^z$ basis which can be performed efficiently based on the Lanczos algorithm and from the NNVB approach which reduces the Hilbert space significantly through the restriction to nearest neighbor VB states. This approach makes it possible to study honeycomb samples of up to $126$ sites using 
space group symmetries\footnote{Although this bipartite quantum dimer model allows for winding number sectors we currently do not need to exploit this additional symmetry.}.

One drawback of both the NNVB and the effective quantum dimer model approach is that they do not presently allow to gauge the quality of the approximation with respect to the
Heisenberg model within the methods themselves. One therefore needs to compare energies or overlaps with exact diagonalization data of the original Heisenberg model for smaller system sizes
in order to locate the regions in the phase diagram where the NNVB approximation is valid.

\section {Comparison Between NNVB and QDM}
\label{sec:comp_NNVB_QDM}

\begin{figure}[!htb]
\begin{center}
\includegraphics [width=\linewidth]{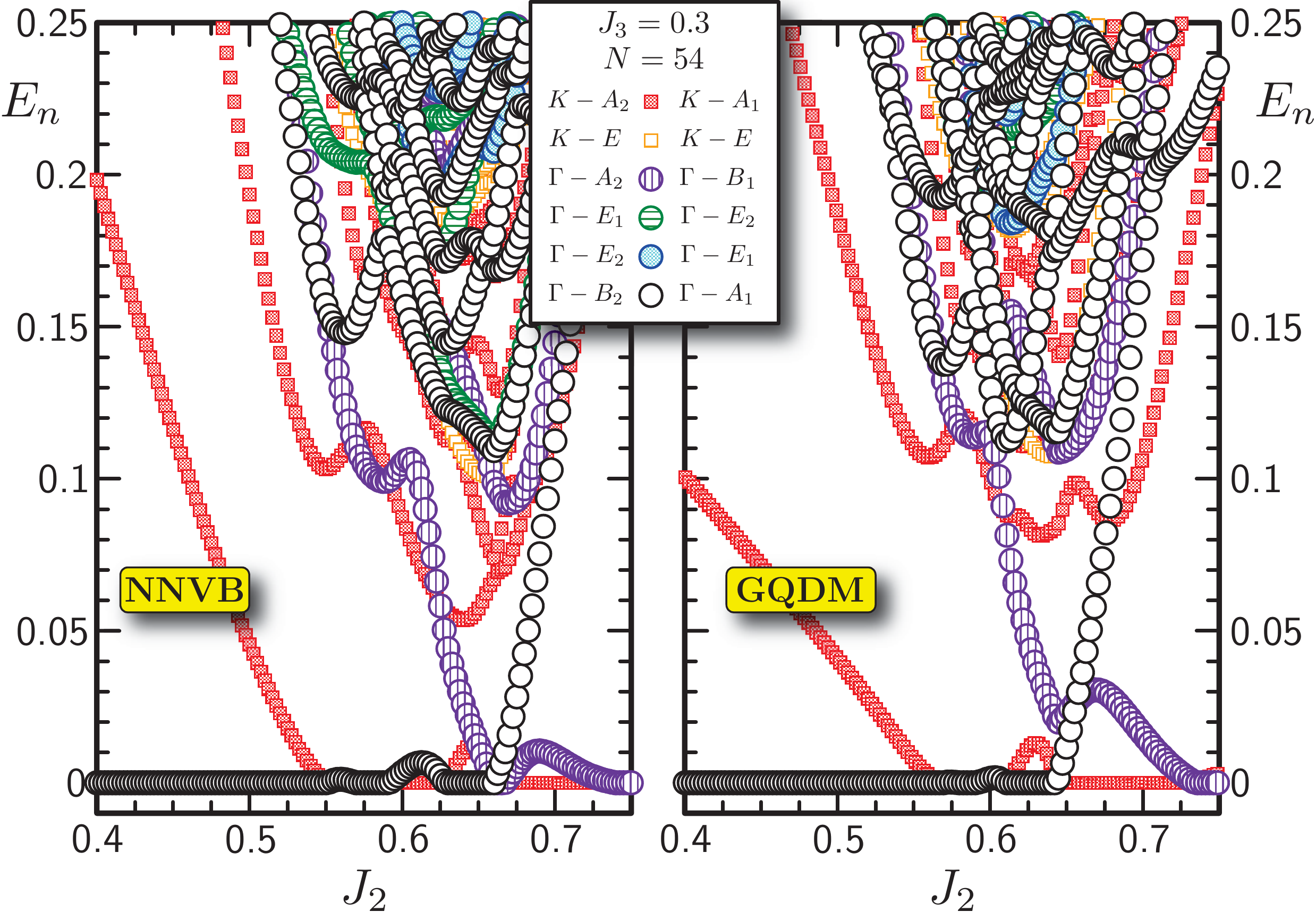}
\caption{(Color online) Low-energy spectra for the frustrated model Eq.~\eqref{eqn:HeisenbergHamiltonian}, for $J_3=0.3$ and as a function of $J_2$,
                obtained by: (a) solving the GEP [Eq.~(\ref{eq:GEP})] in the NNVB subspace and (b) performing EDs for the effective QDM derived in the Appendix
                \ref{sec:derivationeffQDM}.\cite{subtle} In both panels, results have been obtained by diagonalization of an $N=54$ site cluster and energies are
                relative to the ground-state energy. Note that this plot only serves to compare the NNVB and the effective QDM approach on a technical level, because the $J_2$ values
                	considered here are beyond the domain of validity of these approaches for the original Heisenberg model.
\label{fig:SpecN54_VBxQDM}}
\end{center}
\end{figure}
\begin{figure}[!htb]
\begin{center}
\includegraphics [width=\linewidth]{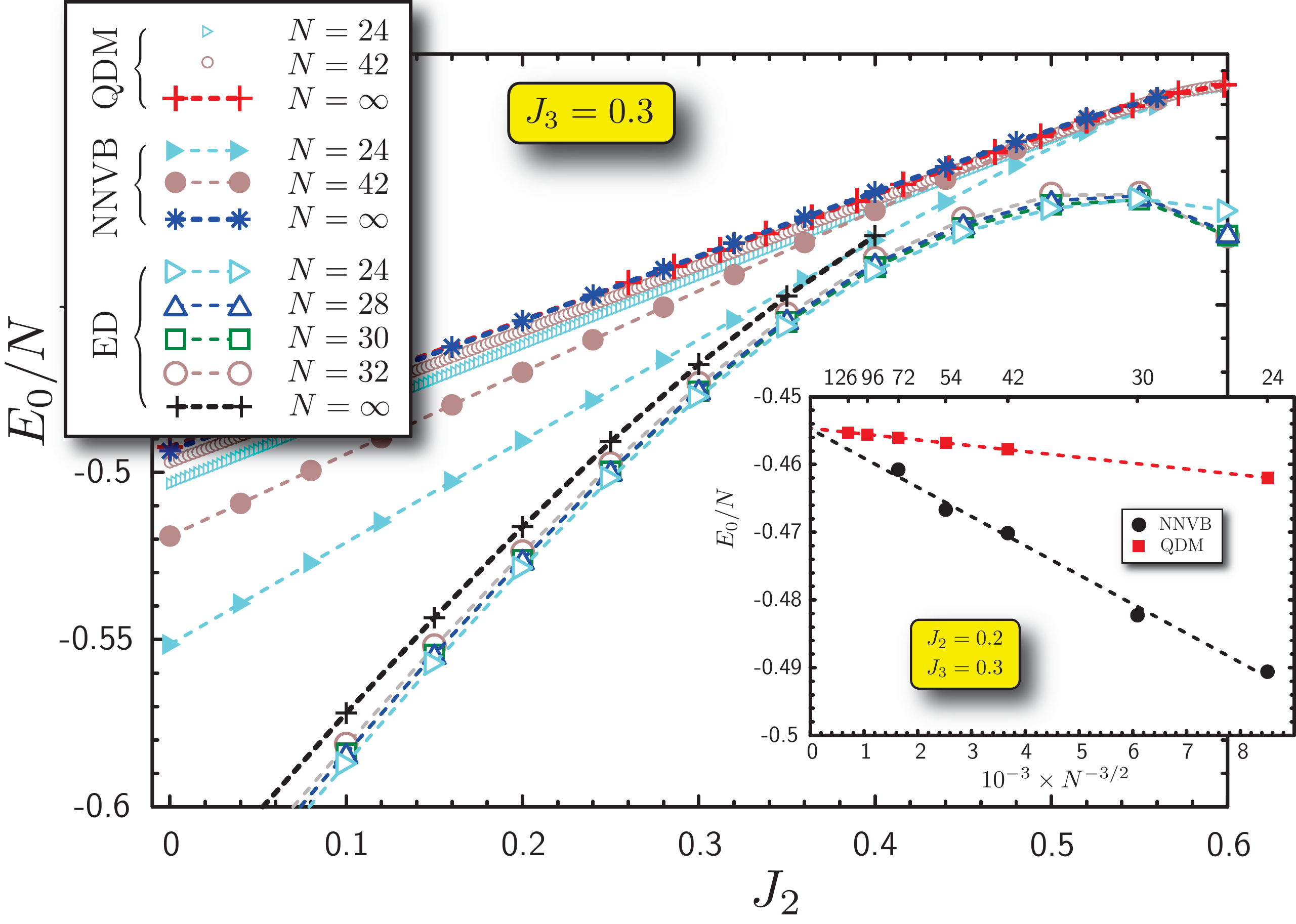}
\caption{(Color online) Ground-state energies for various sizes vs $J_2$ ($J_3=0.3$) obtained by different numerical techniques. Extrapolations to the thermodynamic limit (see text) are also plotted. Inset: 
for $(J_2,J_3)=(0.2,0.3)$, finite-size scaling of the ground-state energy obtained with ED of NNVB and QDM models.
\label{fig:E0_Extrapolation}}
\end{center}
\end{figure}

Although both the NNVB method discussed in Sec.~\ref{subsubsec:NNVB}
and the approach relying on EDs of an effective QDM (Appendix
\ref{sec:derivationeffQDM}) are similar in spirit, for both of them
are formulated in terms of NNVB degrees of freedom and are thus
especially suitable to the study of quantum spin liquids and VBC states, 
they differ somewhat in detail. One such difference concerns the fact that, in
deriving an effective QDM, the diagramatic expansion detailed in
Appendix~\ref{sec:derivationeffQDM} must eventually be truncated, but
one lacks built-in indicators of the convergence of the resulting
expression. On the other hand, overlaps are exactly dealt with within
the NNVB approach (Sec.~\ref{subsubsec:NNVB}), which is therefore
immune to this problem, but this advantage comes at the cost that the
system sizes that can be analyzed via the NNVB approach are
more restricted than those that can be handled by diagonalizing
effective QDMs. Furthermore, and as far as finite-size analysis is
concerned, the formalism described in Appendix
\ref{sec:derivationeffQDM} has the advantage that the amplitudes
appearing in the effective QDM are computed on an infinite
lattice,\cite{Schwandt2010} implying that faster convergence to the
thermodynamic limit is attainable within this approach. Altogether,
these features imply that the formalisms detailed in
Sec.~\ref{subsubsec:NNVB} and Appendix \ref{sec:derivationeffQDM}
should be regarded as complementary to one another. In this sense,
extensive comparisons between the results obtained from both methods
and, due to the variational nature of the NNVB subspace, from
unbiased techniques such as ED are clearly called for.

As a step toward this goal, in Fig.~\ref{fig:SpecN54_VBxQDM} we compare the low energy spectra obtained from NNVB and by diagonalizing the effective QDM derived
in the App.~\ref{sec:derivationeffQDM} for the spin model Eq.~\eqref{eqn:HeisenbergHamiltonian} with $J_3=0.3$ and varying values of $J_2$ (in both cases, an
$N=54$ site cluster has been considered). We first remark that overall features are similar in both spectra, in spite of the subtlety that energy levels displaying similar
dependence on $J_2$ are characterized by different quantum numbers in Fig.~\ref{fig:SpecN54_VBxQDM}(a) and Fig.~\ref{fig:SpecN54_VBxQDM}(b).\cite{subtle}
Another feature salient in Fig.~\ref{fig:SpecN54_VBxQDM} concerns the fact that, by increasing the value of $J_2$, one enters a regime characterized by the occurrence
of successive level crossings. Note that this plot only serves to compare the NNVB and the effective QDM approach on a technical level, because the $J_2$ values
considered here are beyond the domain of validity of these approaches for the original Heisenberg model.

We proceed to a more systematic comparison and in Fig.~\ref{fig:E0_Extrapolation} we plot the ground state energy dependence on $J_2$ for $J_3=0.3$, as obtained
from NNVB and EDs for the effective QDM, for system sizes $N=24$ and $42$ (data for other $N$ are shown only in the inset, but are fully consistent with the analysis
that follows). We first notice that much stronger finite size effects are indeed observed for the NNVB data, in agreement with our discussion above. In extrapolating to
the thermodynamic limit we heuristically assume that the scaling relation $E_{\rm 0}/N \sim N^{-\frac{3}{2}}$, only justified in the case of the N\'eel phase (see
Sec.~\ref{subsec:neel_stability}), also applies in the present case. As shown in the inset in Fig.~\ref{fig:E0_Extrapolation}, this indeed seems to be the case.
Extrapolated values for the ground-state energy computed from NNVB and from the analysis of the effective QDM
are also plotted in Fig.~\ref{fig:E0_Extrapolation}, and from the excellent agreement obtained we conclude that the dominant terms are correctly taken into account
by the truncated expansion detailed in Appendix \ref{sec:derivationeffQDM}. Finally, extrapolated data from both approaches based upon NNVB states are compared
against those from EDs in the $S^z$ basis: we observe that agreement is optimal around the region where {\em plaquette} VBC order is strongest for $J_3=0.3$
(Fig.~\ref{fig:NNVB_dimer_structure_factors}) and where a description based on NNVB states should be at its most accurate level.

\section{Correlations in pure VBC states}
\label{sec:PureStates}

In this Appendix, we compute the expectation values of the 4-spin
correlation function for the four candidate VBC states denoted $\vert
\psi_c \rangle$ (Columnar), $\vert \psi_{st} \rangle$ (Staggered),
$\vert \psi_{sw} \rangle$ (s-wave {\em plaquette}) and $\vert \psi_{dw}
\rangle$ (d-wave {\em plaquette}) in the thermodynamic limit (see
Fig.~\ref{fig:StateDefinitions}). For the {\em plaquette} state indeed, we may consider s-wave or d-wave linear combinations of the two VB coverings of a single hexagon. 
For infinite systems each of these
states is degenerate since it breaks spatial symmetries. This
degeneracy is lifted at finite size and, in order to allow direct
comparison with finite size numerical results, we consider symmetrized
trial states with $(0,0)$ momentum and belonging to the
trivial point group representation A1.

{\it Orthogonality}. The overlap $\langle \psi_\alpha^{i} \vert
\psi_\alpha^{j} \rangle$ between two distinct components of $\vert
\psi_\alpha \rangle$ vanishes exponentially. This point is rather
obvious for $\vert \psi_c \rangle$ and $\vert \psi_{st} \rangle$ but
deserves more attention for $\vert \psi_{sw} \rangle$ and $\vert
\psi_{dw} \rangle$. Generically $\langle \psi_\alpha^{i} \vert
\psi_\alpha^{j} \rangle = 2^{n_l (i,j)-N/2}$ with $N$ the size of the
system and $n_l (i,j)$ the number of loops of the overlap diagram
obtained by superimposing the dimer coverings $i$ and $j$. A direct
inspection of such a diagram shows that $n_l (i,j)=N/6$ for the
{\em columnar} state and $n_l (i,j)=\sqrt{N/2}$ for the {\em staggered} state,
hence showing that any pair of distinct components becomes orthogonal
in the thermodynamic limit.

\begin{figure}[!ht]
\begin{center}
\includegraphics*[width=0.8\linewidth]{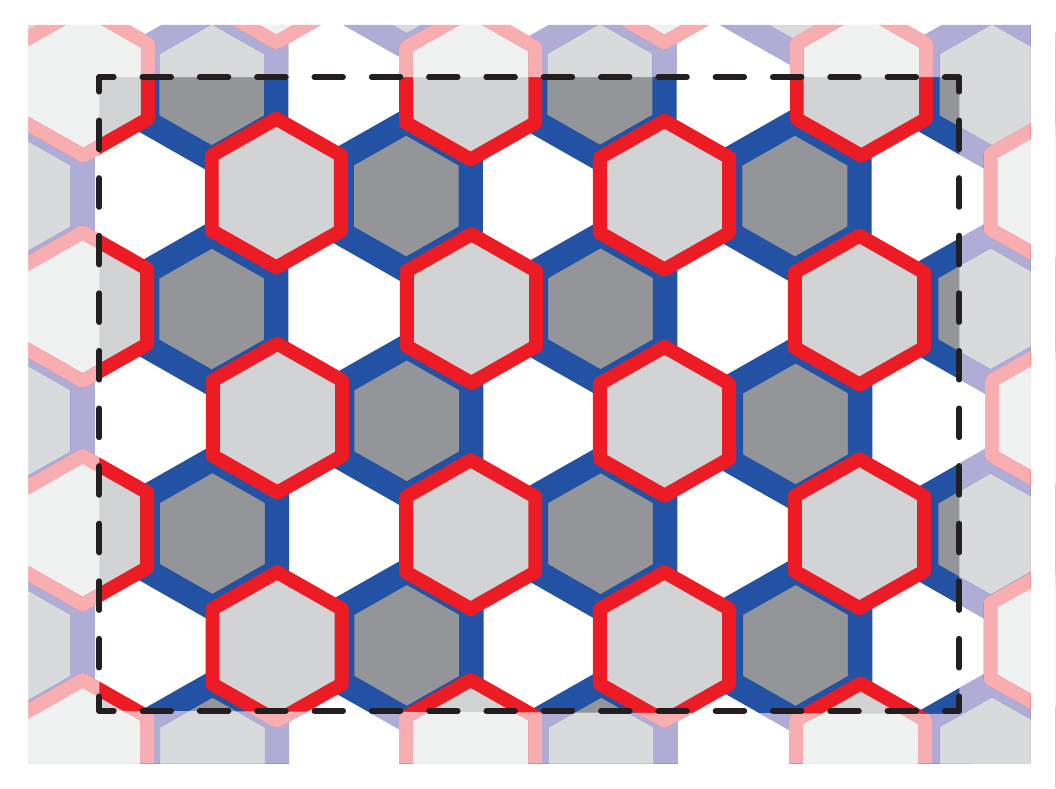}
\caption{Overlap $\langle \psi_\alpha^{i} \vert \psi_\alpha^{j} \rangle$ between two distinct {\em plaquette} state components $\vert \psi_\alpha^{i} \rangle$ (blue) and $\vert \psi_\alpha^{j} \rangle$ (red).}
\label{fig:StateOrthogonality}
\end{center}
\end{figure}

The {\em plaquette} state cases $\alpha = sw$ and $\alpha = dw$ are slightly
more involved since $\langle \psi_\alpha^{i} \vert \psi_\alpha^{j}
\rangle$ includes $2^{N/3}$ overlap contributions (see
Fig.~\ref{fig:StateOrthogonality}). It is possible, albeit not very
illuminating, to find an upper bound of this sum of terms that goes to
zero when the systems size goes to infinity.

In fact, such a result is general: the overlap between two periodic
states $\vert \psi \rangle$ and $\vert \psi' \rangle$ related by a
discrete symmetry $\hat{{\cal S}}$ is either $1$ or $0$ in the
thermodynamic limit. Before actually showing this result let us
mention how it can be anticipated using a physical argument. The two
states being periodic, the structure of the scalar product $\langle
\psi\vert \psi' \rangle$ is itself periodic. It is thus tempting to
infer that $\langle \psi\vert \psi' \rangle \sim \alpha^{N_c}$ in the
thermodynamic limit, where $N_c$ is the number of local patterns
(scaling like the number of sites) and $\alpha$ is related to a {\it
  local} overlap or fidelity. In this case, either $\vert \psi \rangle
= \vert \psi' \rangle$ and $\alpha=1$ or $\vert \psi \rangle \neq
\vert \psi' \rangle$ and $\alpha < 1$ which implies $\langle \psi\vert
\psi' \rangle = 0$ for an infinite system. While qualitatively
correct, the scaling $\langle \psi\vert \psi' \rangle \sim
\alpha^{N_p}$ is actually non-trivial. Indeed, the scalar product
$\langle \psi\vert \psi' \rangle$ {\it does not} generically break
into a product of local disconnected terms but may involve arbitrary
scale resonances.

Let us consider a tensor product state $\vert \psi \rangle = \otimes_c
\vert \varphi_c \rangle$, where the same structure $\vert \varphi_c
\rangle$ defined on a cluster $c$ is repeated on the lattice. The
state $\vert \psi' \rangle$ is related to $\vert \psi \rangle$ by
applying the unitary operator $\hat{\cal S}$. Typically in our case,
$c$ is a hexagon, and $\hat{\cal S}$ is a translation that transforms
a hexagon into a neighboring one. Denoting the density matrix$\hat{\rho} = \vert \psi
\rangle \langle \psi \vert = \otimes_c \hat{\rho}_c$ , the overlap can be written,
\begin{equation}
\label{eq:overlap}
\langle \psi \vert \psi' \rangle = {\text{Tr}} \left ( \hat{\cal S} \hat{\rho}\right ) = {\text{Tr}} \left ( \otimes_c \hat{\cal S} \hat{\rho}_c\right ).
\end{equation} 

But since $\hat{\cal S} \hat{\rho}_c$ and $\hat{\cal S} \hat{\rho}'_c$
do not commute in general, the relation $\langle \psi \vert \psi'
\rangle = \prod_c {\text{Tr}} \left ( \hat{\cal S} \hat{\rho}_c\right
)$ does not hold, which illustrates the point raised previously
according to which $\langle \psi \vert \psi' \rangle$ cannot be
interpreted as the product of local quantities.

However using the H\"older inequality for traces we have for any finite $N$, 
\begin{align*}
 \vert \langle \psi \vert \psi' \rangle \vert &= \left \vert {\text{Tr}} \left ( \otimes_c \hat{\cal S} \hat{\rho}_c\right ) \right \vert \\
& \leq {\text{Tr}} \left \vert  \otimes_c \hat{\cal S} \hat{\rho}_c \right \vert \\
& \leq \prod_c \left ( {\text{Tr}}  \left \vert  \hat{\cal S} \hat{\rho}_c \right \vert^{N_c} \right )^{1/N_c}.
\end{align*}
where $\vert \hat{X} \vert$ denotes $(X^{\dagger} X)^{1/2}$ and $N_c$
is the number of clusters $c$ (scaling linearly with the system size
$N$).

Taking $N$ to infinity, $N_c$ also goes to infinity and
\begin{equation*}
 \vert \langle \psi \vert \psi' \rangle \vert \leq \lim_{N_c \rightarrow \infty} \lambda_0^{N_c} \left ( \left | \hat{\cal S} \hat{\rho}_c\right | \right )
\end{equation*}
where $\lambda_0(\hat{X})$ stands for the maximal eigenvalue of the
positive-semidefinite operator $\hat{X}$. It is then straightforward
to obtain the inequality,
\begin{equation}
\label{eq:overlapupperbound}
 \vert \langle \psi \vert \psi' \rangle \vert \leq  \lim_{N_c \rightarrow \infty} \left |  \langle \varphi_c  | \hat{\cal S} | \varphi_c \rangle \right |^{N_c}
\end{equation}

Two cases can occur : (i) $| \varphi_c \rangle$ is an eigenstate of
$\hat{\cal S}$ in which case $\vert \langle \psi | \psi' \rangle \vert
=1$ or (ii) $| \varphi_c \rangle$ is not invariant under $\cal S$
which implies $\left | \langle \varphi_c | \hat{\cal S} | \varphi_c
  \rangle \right | < 1$ and $\vert \langle \psi \vert \psi' \rangle
\vert = 0$.

\begin{figure*}[!ht]
\begin{center}
\includegraphics[width=0.8\linewidth]{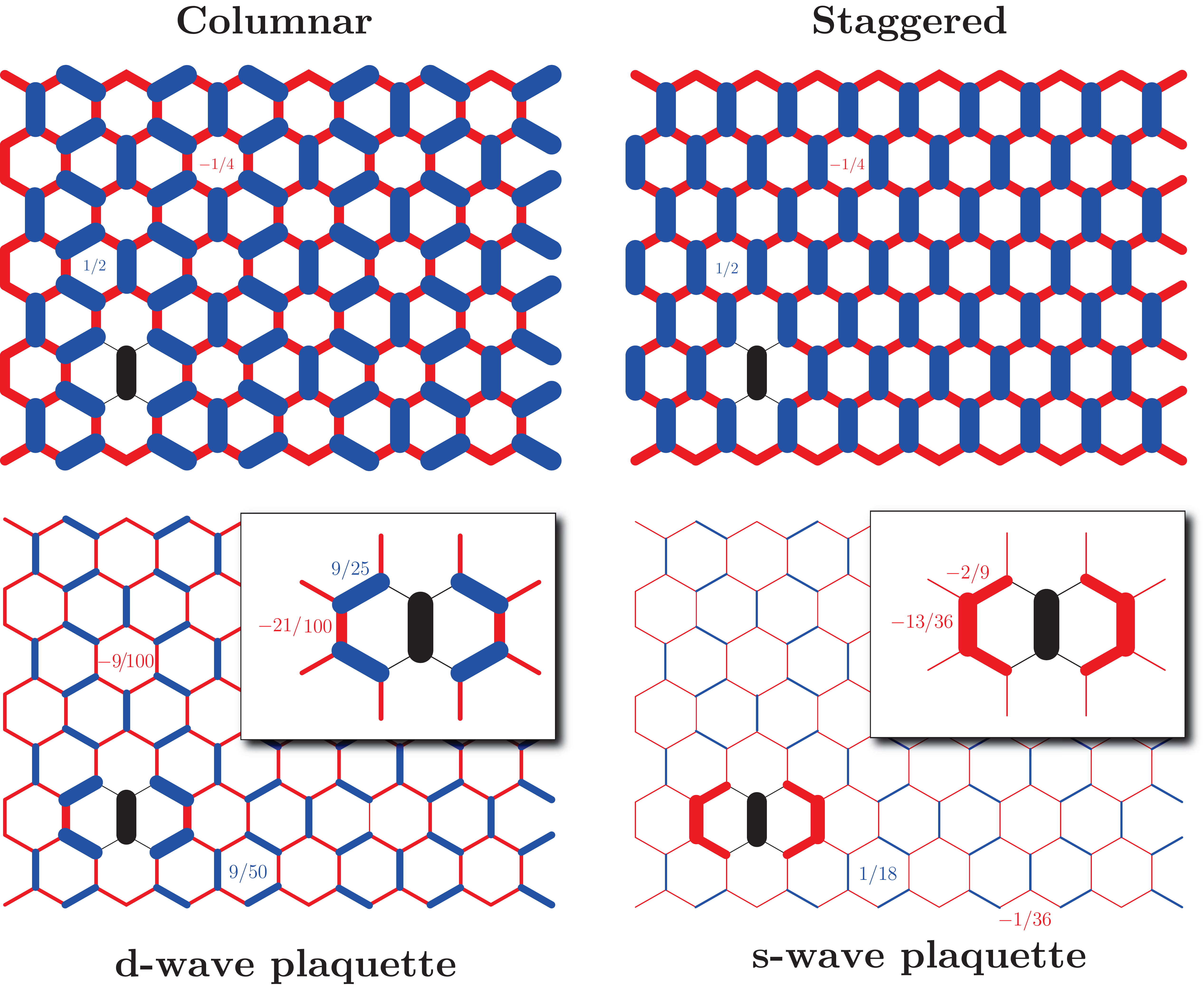}
\caption{4-point correlation function $\langle \hat{P}_{ij} \hat{P}_{kl} \rangle -\langle \hat{P}_{ij} \rangle^2$ in the 4 trial VBC states. The reference bond $(i,j)$ is represented using a thick black line. For plaquette states, the correlations values in the vicinity of the reference bond are different from the bulk values and are displayed in snapshot frames.}
\label{fig:CorrelationSnapshot}
\end{center}
\end{figure*} 
{\it Correlations}. Considering the bond permutation
operators $\hat{P}_{b}$, it is straightforward to remark that $\langle
\psi_\alpha^{i} \vert \hat{P}_{b} \vert \psi_\alpha^{j} \rangle$ and
$\langle \psi_\alpha^{i} \vert \hat{P}_{b} \hat{P}_{b'} \vert
\psi_\alpha^{j} \rangle$ vanish exponentially to $0$ as well since
these operators can only produce local reconfigurations of loops. It
follows that the three components of $\vert \psi_\alpha \rangle$
generate independent contributions to the 4-point correlation function
$\langle \hat{P}_{ij} \hat{P}_{kl} \rangle -\langle \hat{P}_{ij}
\rangle^2$. Its expectation values for the four trial VBC states is
depicted in Fig.~\ref{fig:CorrelationSnapshot}. Note that $\langle
\hat{P}_{ij} \hat{P}_{kl} \rangle -\langle \hat{P}_{ij} \rangle^2 = 4
( \langle (\mathbf{S}_i.\mathbf{S}_j)(\mathbf{S}_k.\mathbf{S}_l)
\rangle -\langle \mathbf{S}_i.\mathbf{S}_j \rangle^2 )$.

Note that in Ref.~\onlinecite{Mosadeq2010}, the authors claim that the three {\em plaquette} states are not orthogonal in the thermodynamic limit, 
which is in contradiction with our general result. However, their approximate numerical values for the dimer-dimer correlations agree with our 
exact ones. In Ref.~\onlinecite{Fouet2001}, the dimer-dimer correlation between parallel bonds on neighboring hexagons is quoted to be 0.01, 
while we find a negative value of $-0.09$, a result which agrees in sign with ED data deep in the {\em plaquette} phase (see Fig.~\ref{fig:FourPoint_N24_many}c).

%


\begin{thebibliography}{95}
\bibitem [{com({\natexlab{a}})}]{comment_scmft_01}%
  \BibitemOpen
  \bibinfo {note} {We notice that within SCMFT not all sites in the $N=8$
  cluster are equivalent and small inhomogeneities ($\lesssim 5\%$ in all cases
  considered) are observed for the mean fields $\langle \mathbf{S}_{j}
  \rangle$.}\BibitemShut {Stop}%
\bibitem [{com({\natexlab{b}})}]{comment_scmft_02}%
  \BibitemOpen
  \bibinfo {note} {SCMFT results for an $N=18$ site cluster for $J_{3}=0$
  display a transition from N\'{e}el to ``disordered" at $J^{\rm c}_{2} \approx
  0.213$. This is to be compared with the transition points $J^{\rm c}_{2}
  \approx 0.131$ and $J^{\rm c}_{2} \approx 0.198$ respectively obtained for
  the $N=6$ and $N=8$ site clusters. This suggests that the location of the
  border for the N\'{e}el phase converges rapidly with $N$ and also constitutes
  further evidence for the existence of non-magnetic phase(s).}\BibitemShut
  {Stop}%
\end{thebibliography}
\end{document}